%% file: ke4_main.tex
\documentclass[11pt]{article}
\usepackage[dvips]{graphicx}
\usepackage{epsfig}

\input{ke4_macros.tex}

\begin{document}
\vspace{15mm}
\centerline{\LARGE EUROPEAN ORGANIZATION FOR NUCLEAR RESEARCH}
\vspace{15mm}
{\flushright{
CERN-PH-EP-2012-185 \\
27 June 2012
 \\}}
\vspace{15mm}

\begin{center}
{\bf {\Large \boldmath{New measurement of the $\ch$ $(\KEQ)$ decay Branching Ratio and Hadronic Form Factors}\unboldmath}}
\end{center}
\vspace{30mm}
\begin{center}
{The NA48/2 Collaboration$\,$\footnotemark[1]}
\end{center}
\input{ke4_abstract.tex}

\noindent
\begin{center}
\it{Submitted for publication in Physics Letters B }
\end{center}

\footnotetext[1]{
\noindent contact: spasimir.balev@cern.ch, brigitte.bloch-devaux@cern.ch\\
\noindent Copyright CERN for the benefit of the NA48/2 Collaboration
}
\renewcommand{\thefootnote}{\arabic{footnote}}
\setcounter{footnote}{0}
\clearpage
\input{ke4_authors.tex}

\clearpage

\section{Introduction}
\input{ke4_intro.tex}

\section{The NA48/2 experiment beam and detector}
\input{ke4_beamdet.tex}

\section{Branching ratio measurement}

\noindent The $\KEQ$ rate is measured relative to the abundant $\KTAU$ 
normalization channel (denoted $\KTP$ below).  
As the topologies of the two modes are similar in terms of number of charged 
particles, the two samples are collected concurrently using the same trigger 
logic and a common event selection is considered as far as possible. This leads
to partial cancellation of the systematic effects induced by an imperfect kaon 
beam description, local detector inefficiencies and a trigger inefficiency. 
The $\KEQ$ rate relative to $\KTP$ and the $\KEQ$ branching ratio (BR) are 
obtained as: 
\begin{eqnarray}\label{eq:rate}
\Gamma(\KEQ) / \Gamma(\KTP) = \frac{N_{s} - N_{b}}{N_{n}} \cdot \frac{A_{n} ~\vep_{n}}{A_{s}~\vep_{s} }
\end{eqnarray}
and
\begin{eqnarray}\label{eq:br}
{\rm BR}(\KEQ) = 
\frac{N_{s} - N_{b}}{N_{n}} \cdot \frac{A_{n} ~\vep_{n}}{A_{s}~\vep_{s} } 
\cdot {\rm BR}(\KTP),
\end{eqnarray}
where $N_{s}, N_{b}, N_{n}$ are the numbers of signal, background and
normalization candidates (the background in the normalization sample is negligible), 
$A_{s}$ and  $\vep_{s}$ are the geometrical
acceptance and trigger efficiency for the signal sample, $A_{n}$ and $\vep_{n}$
are those of the normalization sample.
The normalization branching ratio value BR$($\KTP$) = (5.59 \pm 0.04)\%$ is the
world average as computed in~\cite{pdg}. 

It should be noted that the $\KEQM$ rate has never been measured. As no 
difference is expected from the $\KEQP$ rate~\cite{leewu66}, a comparison of 
the separate measurements of the $\KPL$ and $\KMI$ rates is used as a 
consistency check. 

\subsection{Event selection}
\label{sec:selevt}
\input{ke4_evtsel.tex}

\subsection{Acceptance calculation}
\input{ke4_accep.tex}
\label{sec:accep}

\subsection{Trigger efficiency}

Trigger efficiencies are measured from the data using a minimum bias sample 
downscaled by 100, recorded concurrently with the main analysis data stream.
The control trigger condition for the first level efficiency requires at least
one coincidence of hits in the two planes of the scintillator hodoscope (HOD).
Control triggers for the second level efficiency consist of first level 
triggers recorded regardless of the second level decision.
The overall trigger efficiency is $(98.52 \pm 0.11)\%$ in the signal channel 
and $(97.65 \pm 0.03)\%$ in the normalization channel. 
The observed difference between the two efficiencies can be explained by the
different signal and normalization topologies, four-body and three-body decays 
respectively. Three track events from four-body decays are less 
affected by first level trigger inefficiencies (two tracks in the same HOD 
square region) and by local DCH inefficiencies at the second level 
trigger~\cite{ba07}. The limited statistics of the available control samples 
have a sizable contribution to the systematic error on the $\KEQ$ branching 
ratio measurement.

\subsection{Systematic uncertainties}
\input{ke4_brsyst.tex}
\label{sec:syst}

\subsection{Results}
\input{ke4_brresult.tex}

\section{Form factors normalization measurement}
\subsection{Formalism}
\input{ke4_fftheo.tex}
\input{ke4_ff.tex}

\subsection{Results and discussion}
\input{ke4_ffresult.tex}

\clearpage
\section{Summary}
\input{ke4_conclu.tex}

\section*{Acknowledgments}
We gratefully acknowledge the CERN SPS accelerator and beam line staff for the
excellent performance of the beam and the technical staff of the participating
institutes for their efforts in maintenance and operation of the detector. We
enjoyed fruitful discussions about $\KEQ$ form factors theoretical descriptions
with V.~Bernard, G.~Colangelo, S.~Descotes-Genon, M.~Knecht and P.~Stoffer. 


\end{document}

%% file: ke4_macros.tex
\def\PL#1#2#3{{Phys. Lett. }{\bf B#1 }(#2) #3}

\def\PRV#1#2#3{{Phys. Rev. }{\bf #1 }(#2) #3}
\def\PRD#1#2#3{{Phys. Rev. }{\bf D#1 }(#2) #3}

\def\NPB#1#2#3{{Nucl. Phys. }{\bf B#1 }(#2) #3}
\def\EPJ#1#2#3{{Eur. Phys. J. }{\bf C#1 }(#2) #3}
\def\CPC#1#2#3{{Comp. Phys. Comm. }{\bf #1 }(#2) #3}

\def\NIMA#1#2#3{{Nucl. Inst. Methods }{\bf A#1 }(#2) #3}
\def\ARNS#1#2#3{{Ann. Rev. Nucl. Sci. }{\bf #1 }(#2) #3}

\def\be{\begin{equation}}
\def\ee{\end{equation}}
\def\vep{\varepsilon}
\newcommand{\stat}{\mbox{$\mathrm{_{stat}}$}}
\newcommand{\syst}{\mbox{$\mathrm{_{syst}}$}}

\newcommand{\ext}{\mbox{$\mathrm{_{ext}}$}}
\newcommand{\com}{\mbox{$\mathrm{_{norm}}$}}

\newcommand{\expe}{\mbox{$\mathrm{_{exp}}$}}

\newcommand{\GEVcc}{\mbox{$\mathrm{GeV}/{{\it c}^2}$}}
\newcommand{\GEVc}{\mbox{$\mathrm{GeV}/{{\it c}}$}}
\newcommand{\GEV}{\mbox{$\mathrm{GeV}$}}
\newcommand{\MEVcc}{\mbox{$\mathrm{MeV}/{{\it c}^2}$}}
\newcommand{\MEVc}{\mbox{$\mathrm{MeV}/{\it c}$}}

\newcommand{\PPI}{\mbox{$\pi^{+}$}}
\newcommand{\PMP}{\mbox{$\pi^{\pm}$}}
\newcommand{\MPI}{\mbox{$\pi^{-}$}}
\newcommand{\PK}{\mbox{$p_{K}$}}
\newcommand{\KPM}{\mbox{$K^\pm$}}
\newcommand{\KPL}{\mbox{$K^{+}$}}
\newcommand{\KMI}{\mbox{$K^{-}$}}
\newcommand{\KTP}{\mbox{$K_{3\pi}$}}
\newcommand{\KEQ}{\mbox{$K_{e4}$}}
\newcommand{\KEQP}{\mbox{$K_{e4}^{+}$}}
\newcommand{\KEQM}{\mbox{$K_{e4}^{-}$}}
\newcommand{\KEQPM}{\mbox{$K_{e4}^{\pm}$}}

\newcommand{\KPPD}{\mbox{$K_{\pi\pi^{0}_{D}}$}}

\newcommand{\MEN}{\mbox{$M_{e\nu}$}}
\newcommand{\SE}{\mbox{$S_{e}$}}
\newcommand{\ch}{\mbox{$\KPM \rightarrow \PPI \MPI e^{\pm} \nu$}}
\newcommand{\PDAL}{\mbox{$\pi^{0}_{D} \rightarrow e^{+} e ^{-} \gamma$}}
\newcommand{\PEEG}{\mbox{$\pi^{0} \rightarrow e^{+} e^{-} \gamma$}}

\newcommand{\MES}{\mbox{$m_{e}^2$}}
\newcommand{\ZE}{\mbox{$z_{e}$}}
\newcommand{\TE}{\mbox{$\theta_{e}$}}
\newcommand{\EE}{\mbox{$e^{+} e^{-}$}}
\newcommand{\STE}{\mbox{$\sin\theta_{e}$}}
\newcommand{\STTE}{\mbox{$\sin2\theta_{e}$}}
\newcommand{\CTE}{\mbox{$\cos\theta_{e}$}}
\newcommand{\CTTE}{\mbox{$\cos2\theta_{e}$}}
\newcommand{\SSTE}{\mbox{$\sin^2\theta_{e}$}}

\newcommand{\MPP}{\mbox{$M_{\pi\pi}$}}
\newcommand{\SP}{\mbox{$S_{\pi}$}}
\newcommand{\PIo}{\mbox{$\pi^{0}$}}

\newcommand{\KTAU}{\mbox{$\KPM \rightarrow \PPI \MPI \PMP $}}
\newcommand{\KDA}{\mbox{$\KPM \rightarrow \PMP \PIo $}}
\newcommand{\KDAL}{\mbox{$\KPM \rightarrow \PMP \PIo (\PIo) $}}
\newcommand{\STP}{\mbox{$\sin\theta_{\pi}$}}

\newcommand{\CTP}{\mbox{$\cos\theta_{\pi}$}}

\newcommand{\CPH}{\mbox{$\cos\phi$}}
\newcommand{\SPH}{\mbox{$\sin\phi$}}

\newcommand{\SSTP}{\mbox{$\sin^2\theta_{\pi}$}}

\newcommand{\QRT}{\mbox{$\frac{1}{4}$}}
\newcommand{\HLF}{\mbox{$\frac{1}{2}$}}

%% file: ke4_abstract.tex
\begin{abstract}
A sample of more than one million $\ch$ ($\KEQ$) decay candidates with less 
than one percent background contamination has been collected by the NA48/2 
experiment at the CERN SPS in 2003--2004, allowing a detailed study of 
the decay properties. The branching ratio, inclusive of $\KEQ_\gamma$ decays, 
is measured to be
BR$(\KEQ) = (4.257 \pm 0.016\expe \pm0.031\ext) \times 10^{-5}$
with a total relative error of $0.8\%$.
This measurement complements the study of S- and P-wave hadronic form 
factors by assigning absolute values to the relative hadronic form factors 
obtained earlier in a simultaneous analysis of the $\pi\pi$ scattering 
lengths conducted on the same data sample. The overall form factor 
normalization $f_s = 5.705 \pm 0.017 \expe \pm 0.031 \ext$ is obtained with a
total relative precision of $0.6\%$.
\end{abstract}

%% file: ke4_authors.tex
\begin{center}
{\Large The NA48/2 Collaboration}\\
\vspace{2mm}
 J.R.~Batley,
 G.~Kalmus,
 C.~Lazzeroni$\,$\footnotemark[1],
 D.J.~Munday,
 M.W.~Slater$\,$\footnotemark[1],
 S.A.~Wotton \\
{\em \small Cavendish Laboratory, University of Cambridge,
Cambridge, CB3 0HE, UK$\,$\footnotemark[2]} \\[0.2cm]
 R.~Arcidiacono$\,$\footnotemark[3],
 G.~Bocquet,
 N.~Cabibbo$\,$\footnotemark[4],
 A.~Ceccucci,
 D.~Cundy$\,$\footnotemark[5],
 V.~Falaleev,
 M.~Fidecaro,
 L.~Gatignon,
 A.~Gonidec,
 W.~Kubischta,
 A.~Norton$\,$\footnotemark[6],
 A.~Maier,
 M.~Patel$\,$\footnotemark[7],
 A.~Peters\\
{\em \small CERN, CH-1211 Gen\`eve 23, Switzerland}
% ...
\\[0.5cm]
\end{center}

\setcounter{footnote}{0}
\footnotetext[1]{University of Birmingham, Edgbaston, Birmingham,
B15 2TT, UK}
\footnotetext[2]{Funded by the UK Particle Physics and Astronomy
Research Council}
\footnotetext[3]{Universit\`a degli Studi del Piemonte Orientale e Sezione 
dell'INFN di Torino, I-10125 Torino, Italy}
\footnotetext[4]{Deceased}
\footnotetext[5]{Istituto di Cosmogeofisica del CNR di Torino,
I-10133 Torino, Italy}
\footnotetext[6]{Dipartimento di Fisica dell'Universit\`a e Sezione
dell'INFN di Ferrara, I-44100 Ferrara, Italy}
\footnotetext[7]{Department of Physics, Imperial College, London,
SW7 2BW, UK}
\footnotetext[8]{CERN, CH-1211 Gen\`eve 23, Switzerland}
\footnotetext[9]{Physikalisches Institut, Ruprecht-Karls-Universit\"at Heidelberg, D-69120 Heidelberg, Germany}
\footnotetext[10]{Faculty of Physics, University of Sofia ``St. Kl.
Ohridski'', 
 1164 Sofia, Bulgaria, funded by the Bulgarian National Science Fund under contract DID02-22}
\footnotetext[11]{Northwestern University,
Evanston, IL 60208, USA}
\footnotetext[12]{Centre de Physique des Particules de Marseille,
IN2P3-CNRS, Universit\'e de la M\'editerran\'ee, F-13288 Marseille,
France}
\footnotetext[13]{Department of Physics and Astronomy, George Mason
University, Fairfax, VA 22030, USA}
\footnotetext[14]{CP3, Universit\'e Catholique de Louvain, B-1348 
Louvain-la-Neuve, Belgium}
\footnotetext[15]{Dipartimento di Fisica, Universit\`a di Modena e
Reggio Emilia, I-41125 Modena, Italy}
\footnotetext[16]{Istituto di Fisica, Universit\`a di Urbino,
I-61029 Urbino, Italy}
\footnotetext[17]{Helmholtz-Institut Mainz, Universit\"at Mainz,
D-55099 Mainz, Germany}
\footnotetext[18]{Funded by the German Federal Minister for
Education and research under contract 05HK1UM1/1}
\footnotetext[19]{SLAC, Stanford University, Menlo Park, CA 94025,
USA}
\footnotetext[20]{Laboratory for High Energy Physics, CH-3012 Bern,
Switzerland}
\footnotetext[21]{UCLA, Los Angeles, CA 90024, USA}
\footnotetext[22]{Laboratori Nazionali di Frascati, via E. Fermi,
40, I-00044 Frascati (Rome), Italy}
\footnotetext[23]{Dipartimento di Fisica dell'Universit\`a e Sezione dell'INFN di Padova, I-35131 Padova, Italy}
\footnotetext[24]{Instituto de F\'{\i}sica Corpuscular IFIC,
Universitat de Val\`{e}ncia, E-46071 Val\`{e}ncia, Spain}
\footnotetext[25]{Dipartimento di Fisica Sperimentale
dell'Universit\`a di Torino, I-10125 Torino, Italy}
\footnotetext[26]{Institut de Physique Nucleaire de Lyon,
IN2P3-CNRS, Universite Lyon I, F-69622 Villeurbanne, France}
\footnotetext[27]{Funded by the German Federal Minister for Research
and Technology (BMBF) under contract 056SI74}
\footnotetext[28]{University College Dublin School of Physics,
Belfield, Dublin 4, Ireland}
\footnotetext[29]{Centro de Investigaciones Energeticas
Medioambientales y Tecnologicas, E-28040 Madrid, Spain}
\footnotetext[30]{Funded by the Austrian Ministry for Traffic and
Research under the contract GZ 616.360/2-IV GZ 616.363/2-VIII, and
by the Fonds f\"ur Wissenschaft und Forschung FWF Nr.~P08929-PHY}


%% file: ke4_intro.tex
The interest of $\ch$ decays (denoted $\KEQ$ in the following) was recognized 
many years ago at a time when only a handful of such events had been 
observed~\cite{leewu66}. The accumulation of a large sample of more than one 
million of such decays by the NA48/2 
experiment has recently allowed a very detailed study of the $\pi \pi$ 
scattering lengths and hadronic form factors~\cite{ke410}. In that study, the
$I = 0$ and $I = 2$ S-wave scattering lengths have been determined with an 
improved precision comparable to the few percent relative accuracy of the most 
elaborate theoretical predictions~\cite{chpt}. Without the branching ratio 
value, only relative form factors could be measured, giving a full set 
of values up to a common normalization. 

A new measurement of the $\KEQP$ and $\KEQM$ decay rates based on the data
collected by the NA48/2 experiment at the CERN SPS in 2003--2004 is
reported here. The event sample is about three times larger than the total 
world sample and has one percent level background contamination. A good 
control of systematic
uncertainties, dominated by the external error from the normalization 
mode, allows rate and form factors to be measured with an improved precision.
These can be used as input to the determination of the Low 
Energy Constants (LEC) of Chiral Perturbation Theory 
(ChPT)~\cite{ke4ff1,ke4ff2,ke4ff3} and as tests of other theoretical dispersive
approaches~\cite{ke4disp}.

%% file: ke4_beamdet.tex
The NA48/2 experiment, specifically designed for charge asymmetry
measurements~\cite{ba07}, takes advantage of simultaneous $\KPL$ and $\KMI$ 
beams produced by $400~\GEVc$ primary CERN SPS protons impinging on a 40~cm 
long beryllium target. Oppositely charged particles, with a central momentum of
$60~\GEVc$ and a momentum band of $\pm 3.8\%$ (rms), are selected by 
two systems of dipole magnets with zero total deflection (each system forming 
an `achromat'), focusing quadrupoles, muon sweepers and collimators.

At the entrance of the decay volume housed in a 114 m long evacuated
vacuum tank, the beams contain $\sim3.6\times10^6$ charged kaons
per pulse of about 4.5 s duration with a flux ratio
$\KPL / \KMI$ close to 1.8. Both beams follow the same path in the decay 
volume: their axes coincide within 1~mm, while the transverse size of each beam
is about 1~cm. 

The decay volume is followed by a magnetic spectrometer 
located in a tank filled with helium at nearly atmospheric pressure, separated 
from the vacuum tank by a thin 
($0.3\%X_0$) $\rm{Kevlar}\textsuperscript{\textregistered}$ window. 
An aluminum beam pipe of 16~cm outer diameter
traversing the centre of the spectrometer (and all the following
detector elements) allows the undecayed beam particles and the muon
halo from decays of beam pions to continue their path in vacuum. The
spectrometer consists of four octagonal drift chambers (DCH), each composed of 
four staggered double planes of sense wires, and located upstream (DCH1--2) and
downstream (DCH3--4) of a large aperture dipole magnet. 
The magnet provides a transverse momentum kick $\Delta p=120~\MEVc$ to 
charged particles in the horizontal plane. 
The momentum resolution achieved in the 
spectrometer is $\sigma_p/p = (1.02 \oplus 0.044\cdot p)\%$ ($p$ in $\GEVc$).

The spectrometer is followed by a hodoscope (HOD) consisting of two planes of 
plastic scintillator segmented into vertical and horizontal strip-shaped 
counters (128 in total). The HOD surface is logically subdivided into 16 
exclusive square regions whose fast signals are used to trigger the detector 
readout on charged track topologies. Its time resolution is $\sim 150$ ps.

A liquid krypton electromagnetic calorimeter (LKr), used for particle 
identification in the present analysis, is located behind the HOD. It
is an almost homogeneous ionization chamber with an active volume of
7 m$^3$ of liquid krypton, segmented transversally into 13248 projective cells,
approximately 2$\times$2 cm$^2$ each, $27X_0$ deep and with no longitudinal 
segmentation. 
The energies of electrons and photons are measured with a resolution 
$\sigma_E /E = (3.2/\sqrt{E} \oplus 9.0/E \oplus 0.42)\%$ ($E$ in $\GEV$) and
the transverse position of isolated showers is measured with a
spatial resolution $\sigma_x = \sigma_y = (0.42/\sqrt{E} \oplus 0.06)$~cm. 

The muon veto counter (MUV) is located further downstream. It is composed
of three planes of plastic scintillator slabs (aligned horizontally
in the first and last planes, and vertically in the middle plane)
read out by photomultipliers at both ends, each preceded by a 0.8~m thick iron
absorber. The MUV is also preceded by a hadronic calorimeter (not used in
this analysis) with a total iron thickness of 1.2~m.

A more detailed description of the NA48 detector and its performances can be 
found in~\cite{fa07}. 

A dedicated two-level trigger selects and flags the events. At the first level 
(L1), charged track topologies are selected by requiring coincidences of hits 
in the two HOD planes in at least two of the 16 square regions. At the second 
level (L2), a farm of asynchronous microprocessors performs a fast 
reconstruction of tracks and runs a decision-taking algorithm.
This trigger logic ensures a very high trigger efficiency for three-track 
topologies. Inefficiencies are typically a few
$10^{-3}$ at the first level and a few $10^{-2}$ at the second level (more
details can be found in~\cite{ke410,ba07}).

%% file: ke4_evtsel.tex
The same data sample has been considered in both signal and normalization 
studies. Given that ${\rm BR}(\KTP)/{\rm BR}(\KEQ)\simeq 1400$, a 
filtering of the data stream and the analysis are performed in such a way that 
the $\KTP$ candidates are effectively prescaled by a factor of 100 with a 
negligible error, while the $\KEQ$ candidates are not affected, leading to a
significant reduction of the data volume. 

The analysis of the $\pi \pi$ scattering lengths and form factors presented
in~\cite{ke410} focuses on a sample free of hard radiative events at the
price of some cuts on the additional photon activity in the LKr calorimeter. 
The present analysis includes radiative events and thus loosens or
removes some of the selection cuts which could bias the event counting because 
of imperfect modeling of the photon emission mechanism.

\vspace{2mm}
\noindent {\bf Common selection}
\vspace{2mm}

\noindent Three-track vertices (compatible with either $\KEQ$ or $\KTP$ decay 
topology), in events satisfying the three-track trigger logic conditions, are 
reconstructed by backward extrapolation of track segments from the spectrometer
into the decay volume, taking into account the measured stray magnetic field in
the vacuum tank and multiple scattering.
The reconstructed vertex must satisfy the following criteria:\\
--- total charge of the three tracks (called ``vertex tracks'' in the following) equal to $\pm1$; \\
--- longitudinal position of the vertex within the fiducial decay volume, 2 to 
95 m downstream of the final collimator, and its transverse position within 5 
cm of the nominal beam axis; \\
--- vertex tracks consistent in time within 12~ns; no additional in-time tracks
present in the reconstructed event (see section~\ref{sec:syst} for details); \\
--- all vertex tracks within the DCH, HOD, LKr and MUV geometric acceptances;
distance between any track and the beam mean position (monitored with $\KTP$
decays) in the DCH1 plane greater than 12~cm for better time-dependent 
acceptance control; \\
--- track separations required in the DCH1 and LKr planes (minimum allowed
distance 2~cm and 20~cm respectively) to suppress photon conversions and to 
ensure efficient particle identification, minimizing shower overlaps; \\
--- distance from the impact point of each vertex track on the LKr plane to the
closest inactive cell of the calorimeter larger than 2~cm to provide maximum
collection of energy deposit;\\
--- total momentum of the three tracks $|\sum\vec p_i|$ below $70~\GEVc$; \\
--- no track-associated signal allowed in at least two planes of the MUV 
in-time with any vertex track (within 10~ns). \\
If several vertices satisfy the above conditions, the one with the lowest fit 
$\chi^2$ is considered. 

\vspace{2mm}
\noindent {\bf Particle identification}
\vspace{2mm}

\noindent Particle identification criteria are based on the geometric 
association of an in-time LKr energy deposition cluster to a track extrapolated
to the calorimeter front face (denoted ``associated cluster'' below). The ratio
of energy deposition in the LKr calorimeter to momentum measured by the 
spectrometer ($E/p$) is used for pion/electron separation. A track is
identified as an {\it electron} ($e^{\pm}$) if its momentum is greater than
$2.75~\GEVc$ and it has an associated cluster with $E/p$ between 0.9 and 1.1. 
A track is identified as a {\it pion} ($\pi^{\pm}$) if its momentum is above 
$5~\GEVc$ and it has either no associated cluster or an 
associated cluster with $E/p$ smaller than 0.8. 

Powerful further suppression of pions mis-identified as electrons within the 
above conditions is obtained by using a discriminant variable which is a 
linear combination of quantities related to shower properties ($E/p$, radial 
shower width and energy weighted track-cluster distance at LKr front face), 
and is almost momentum independent. The discriminant variable was trained on 
dedicated track samples to be close to 1 for electron tracks and close to 0 for
pion tracks faking electron tracks (the discriminant variable performances are 
shown in section~\ref{sec:syst}). In the signal selection, its value is 
required to be larger than 0.9 for the {\it electron} track.

\vspace{2mm}
\noindent {\bf Signal sample}
\vspace{2mm}

\noindent The $\KEQ$ candidates are then selected using the following particle 
identification and kinematic criteria: \\
--- the vertex is composed of a single {\it electron} candidate and a 
pair of oppositely charged {\it pion} candidates $\pi^+ \pi^-$; \\
--- the invariant mass of the three tracks in the $\pi^+ \pi^- \pi^\pm$ 
hypothesis $(M_{3\pi})$ and the transverse momentum $p_t$ relative to the beam 
axis are outside a half-ellipse centered on the nominal kaon mass~\cite{pdg} 
and zero $p_t$, with semi-axes of $20~\MEVcc$ and $35~\MEVc$, respectively, 
thus requiring a non-zero $p_t$ value for the undetected neutrino and rejecting
fully reconstructed three-body $\KTP$ decays (the $\KEQ$ signal loss from this 
cut is $\sim 4.5 \%$, as shown by simulation); \\
--- the square invariant mass $M_X ^2$ in the $\KPM \rightarrow \PMP X$ decay
is larger than $0.04 ~(\GEVcc)^2$ to reject $\KDA$ decays with a 
subsequent $\PEEG$ decay; \\
--- the invariant mass of the $\EE$ system (assigning an
electron mass to the oppositely charged pion candidate) is larger than 
$0.03~\GEVcc$ to ensure rejection of converted photons and of some multi-$\PIo$
events (as $\KPM \rightarrow \PMP \PIo \PIo $).

Extra rejection of three-body decays is obtained by 
reconstructing the kaon momentum under the assumption of a four-body decay with
an undetected massless neutrino. Imposing energy-momentum conservation in the 
decay and fixing the kaon mass and the beam direction to their nominal values,
a quadratic equation in the kaon momentum $\PK$ is obtained.
A $\KEQ$ candidate is accepted if a solution is found in the nominal range 
between 54 
and 66 $\GEVc$, allowing a small fraction of solutions with negative but close
to zero equation discriminant values as observed for reconstructed simulated 
signal events because of non-perfect resolution
(in this case, a single solution is obtained by setting the equation 
discriminant to zero).

\begin{figure}[htp]
\begin{center}
\begin{picture}(160,150)
\put(-125,-45){\epsfxsize75mm\epsfbox{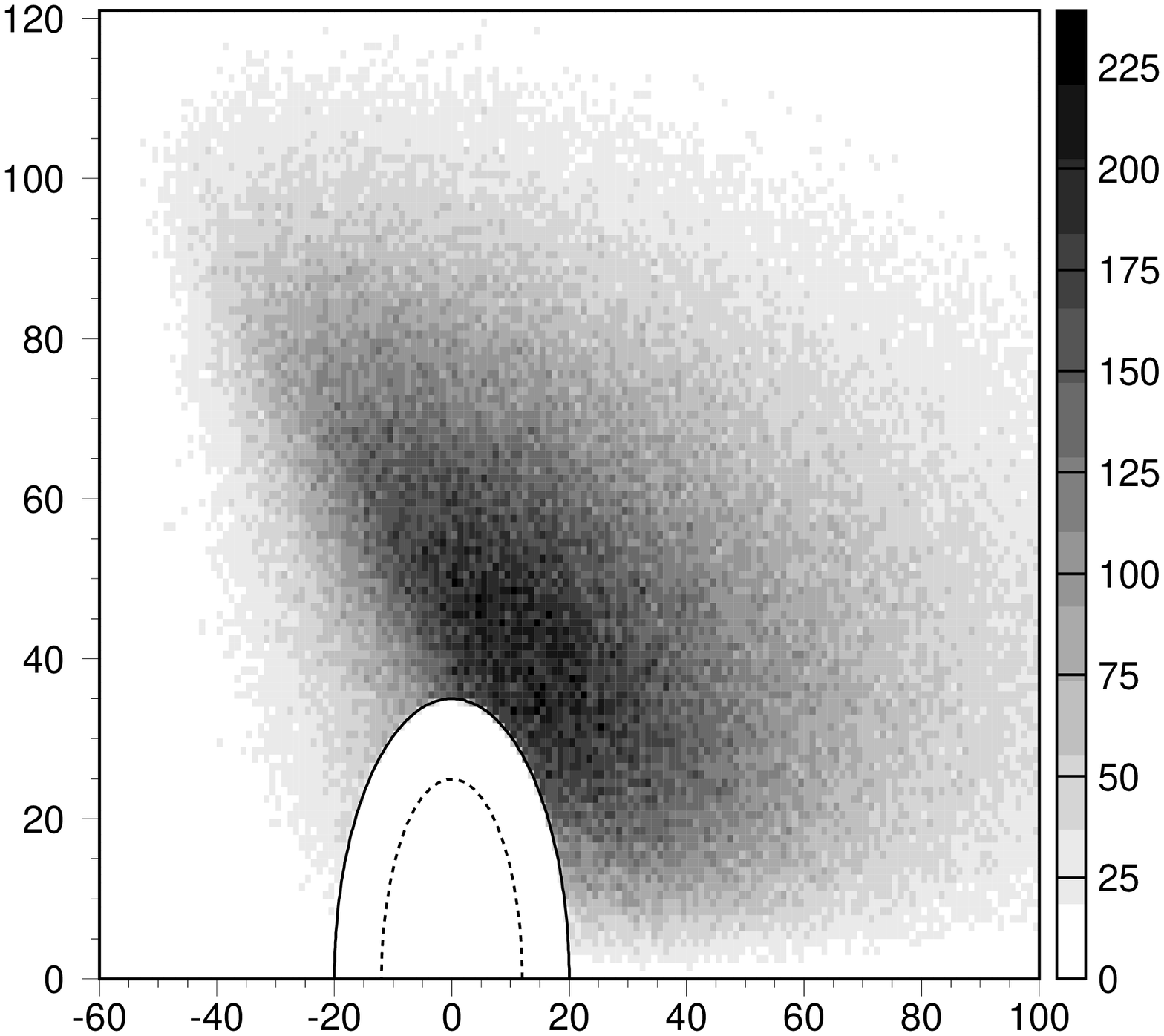}}
\put(100,-45){\epsfxsize75mm\epsfbox{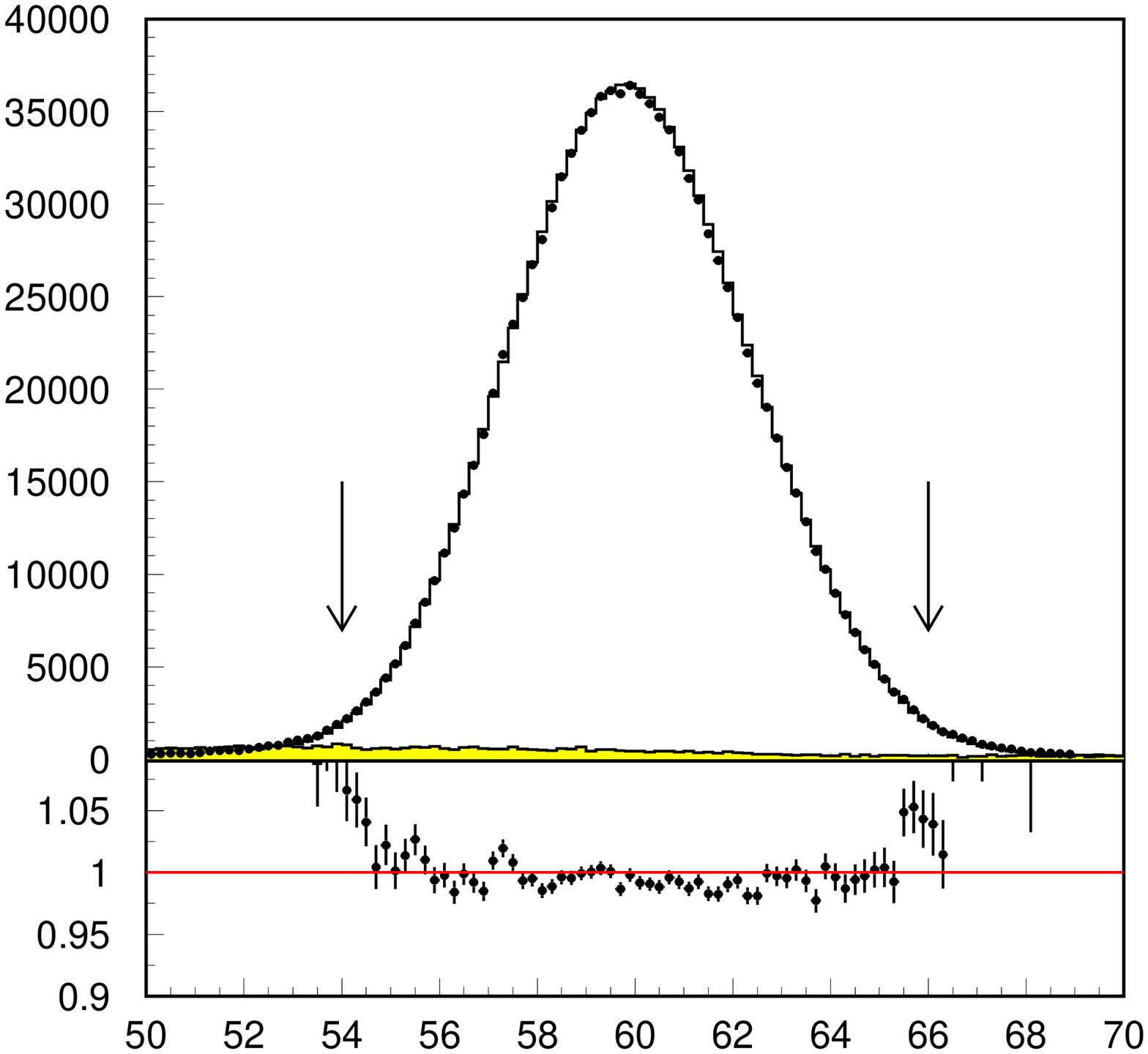}}
\put(200,10){\epsfxsize70mm\epsfbox{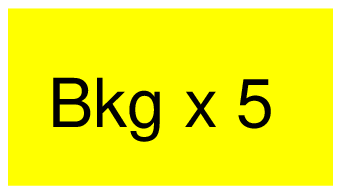}}
\put(-125,160){$p_t  ~~(\MEVc)$}
\put(-35,-50){$M_{3\pi} - M_K ~~(\MEVcc)$}
\put(225,-50){$\PK ~(\GEVc)$}
\put(40,135){\bf \large (a)}
\put(275,135){\bf \large (b)}
\end{picture}
\end{center}
\vspace{5mm} 
\caption{{\bf (a)} Reconstructed ($M_{3\pi},p_t$) plane for the $\KEQ$ signal 
candidates. The elliptic cuts used in the signal (solid line) and normalization
(dashed line) selections are shown. 
{\bf (b)} Reconstructed kaon momentum from signal events for
data after background subtraction (symbols), simulation normalized to data 
(histogram) and background events (scaled by a factor of 5 to be 
visible) as shaded area. The arrows point to the kaon momentum cuts values. 
The lower plot is the ratio of the two spectra (data/simulation) displayed in 
the upper plot.
\label{fig:signal}}
\end{figure}

A total sample of 1 108 941 $\KEQ$ candidates (712 288 $\KPL$ and 396 653 
$\KMI$) were selected from a total of $\sim2.5\times10^{10}$ triggers recorded 
in 2003--2004. The selection is illustrated in Fig.~\ref{fig:signal}a in the 
$(M_{3\pi},p_t)$ plane and the reconstruction in Fig.~\ref{fig:signal}b by the
kaon momentum distribution.

\vspace{5mm}

\vspace{2mm}
\noindent {\bf Background estimate}
\vspace{2mm}

\noindent The $\KTP$ decay is the most significant background source. It
contributes either via the decay in flight of a single pion 
($\pi^\pm \to e^\pm\nu$) or mis-identification of a pion as an electron. 
Only pion decays occurring close to the kaon decay vertex or leading to a 
forward electron and thus consistent with a three-track vertex and satisfying 
the ($M_{3\pi}, p_t$) requirements contribute to the background. Other 
background sources come from $\KDAL$ decays with subsequent Dalitz decay of a 
$\PIo$ ($\PDAL$), an electron mis-identified as a pion, and photon(s) 
undetected. Such two- or three-body decay topologies are very unfavored by the 
signal selection criteria and contribute at sub per mil level.

Decays violating the $\Delta S = \Delta Q$ rule would appear as 
``wrong sign electron'' (WS) $\pi^\pm \pi^\pm e^\mp \nu$ $\KEQ$ candidates and
are expected at a negligible rate (BR $< 1.2 \times 10^{-8}$ at $90\%$ 
confidence level~\cite{pdg}).
The kinematic distribution of the background events is then to a good
approximation identical to that of the reconstructed WS candidates multiplied 
by a factor of 2 as two pions from $\KTP$ decays can mimic the signal final 
state while one pion only contributes to the WS topology. The uncertainty on 
this factor of 2 is discussed in section~\ref{sec:syst}.

\begin{figure}[htp]
\begin{center}
\begin{picture}(160,150)
\put(-125,-55){\epsfxsize75mm\epsfbox{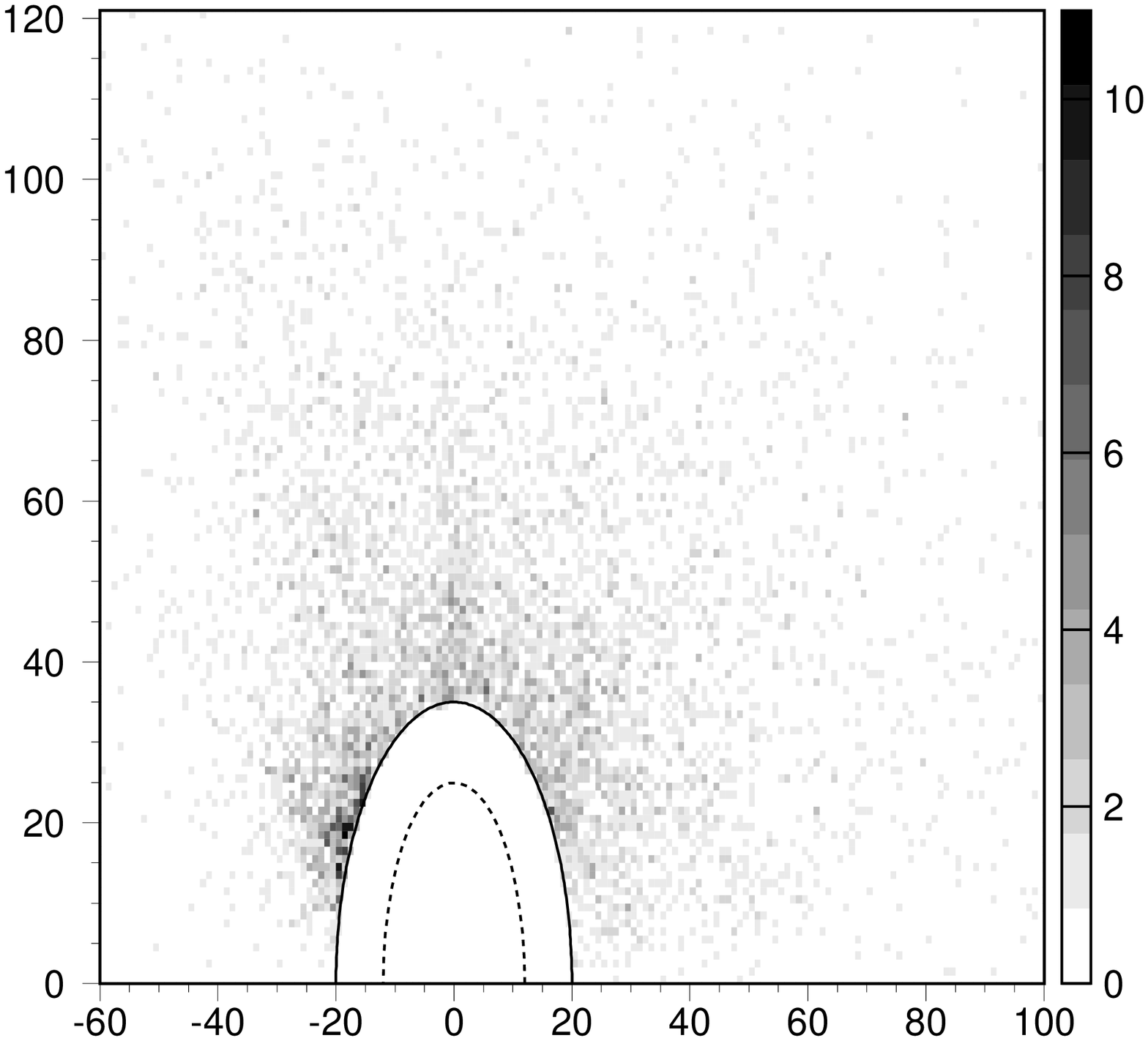}}
\put(100,-55){\epsfxsize75mm\epsfbox{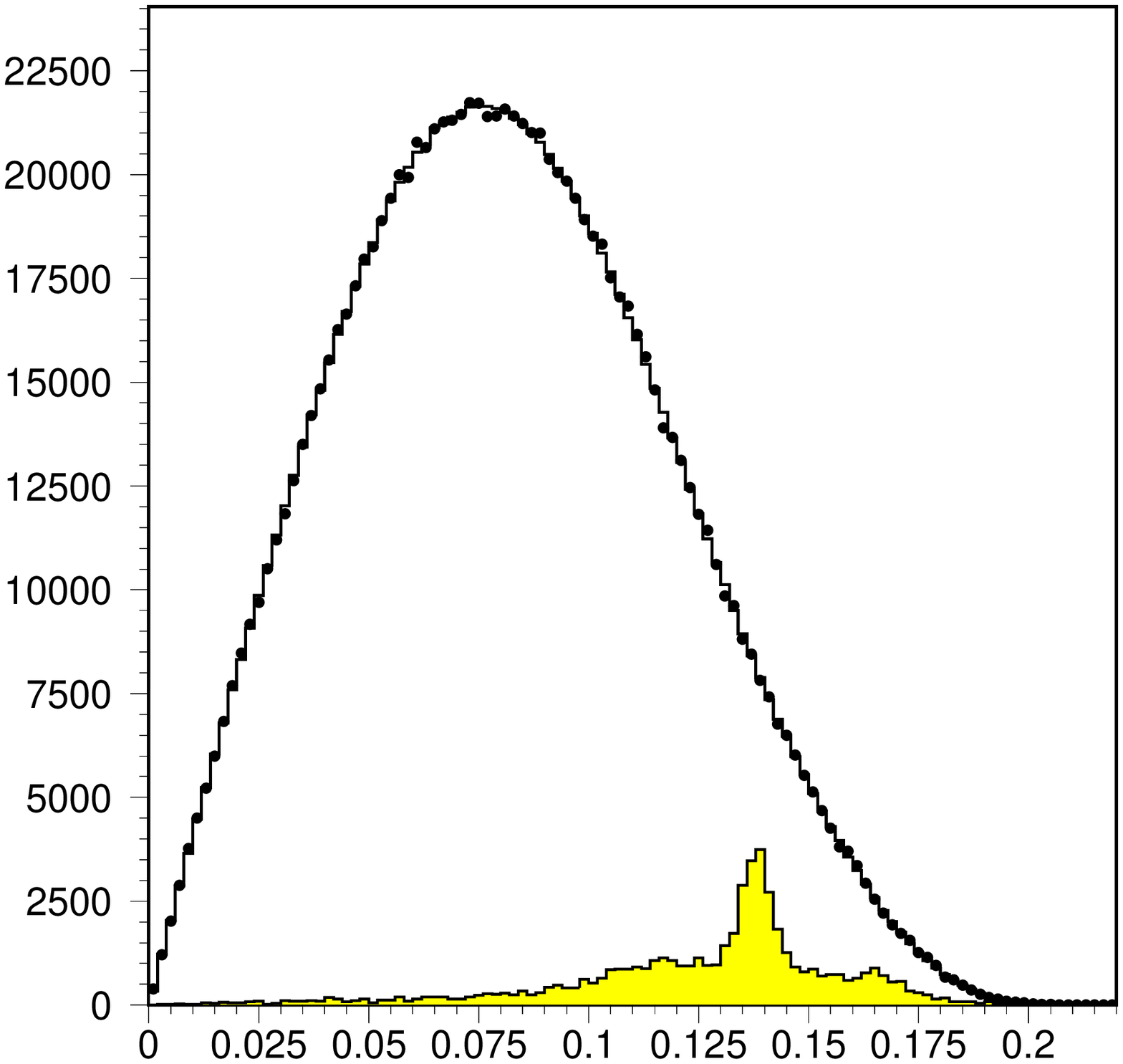}}
\put(200,10){\epsfxsize70mm\epsfbox{figbox5.eps}}
\put(-125,150){\bf  $p_t  ~~(\MEVc)$}
\put(-35,-60){\bf $M_{3\pi} - M_K ~~(\MEVcc)$}
\put(225,-60){\bf $\MEN ~~(\GEVcc)$}
\put(40,125){\bf \large (a) }
\put(270,125){\bf \large (b)}
\end{picture}
\end{center}
\vspace{8mm} 
\caption{{\bf (a)} Reconstructed ($M_{3\pi},p_t$) plane for the 
$\KEQ$ background estimated from WS events. The elliptic cuts used in the 
WS selection (solid line) and normalization (dashed line) selections are shown.
{\bf (b)} Reconstructed dilepton invariant mass for $\KEQ$ events. Data are 
shown as symbols, simulation as histogram  and background events (scaled by a 
factor of 5 to be visible) as shaded area.
\label{fig:WS}}
\end{figure}

Changing the requirement of a pair of opposite charge pions 
($\pi^+ \pi^-$ candidates) in the vertex selection  to a pair of same charge 
pions ($\pi^\pm \pi^\pm$ candidates) and keeping all other requirements 
unchanged is sufficient to determine the number of events in the WS sample.
The distribution of the WS $\KEQ$ candidates in the $(M_{3\pi}, p_t )$ plane 
is displayed in Fig.~\ref{fig:WS}a. Another feature of the WS sample is shown 
in Fig.~\ref{fig:WS}b which displays the reconstructed invariant mass of the 
dilepton system in the signal and WS selections. A peak at the $m_{\pi^+}$ 
value can be seen as expected from $\KTP$ decays followed by a pion decay in 
flight.
 
A sample of 5276 $\KEQ$ WS candidates (3276 $\KPL$ and 2000 $\KMI$) has been
selected concurrently with the signal sample. 
As $\KTP$ decays are the dominant contributors, the total background is then 
estimated to be $2 \times 5276 $ events, a $0.95\%$ relative contamination to 
the signal. The systematic uncertainty on this quantity is discussed in 
section~\ref{sec:syst}.

\vspace{2mm}
\noindent {\bf Normalization sample}
\vspace{2mm}

\noindent The $\KTP$ sample is selected applying the following requirements to
events passing the common selection: \\
--- the vertex is required to be composed of three {\it pion}
$\pi^+ \pi^- \pi^\pm$ candidates; \\
--- the invariant mass of the three tracks in the $\pi^+ \pi^- \pi^\pm $ 
hypothesis $(M_{3\pi})$ and the transverse momentum $p_t$ are inside a 
half-ellipse (as drawn in Fig.~\ref{fig:signal}a) centered on the kaon mass and
zero $p_t$, with semi-axes $12~\MEVcc$ and $25~\MEVc$, respectively, thus 
requiring fully reconstructed $\KTP$ three-body decays; \\
--- the total momentum of the three tracks $|\sum\vec p_i|$ is between 54 and 
66 $\GEVc$.

\begin{figure}[htp]
\begin{center}
\begin{picture}(160,150)
\put(-125,-50){\epsfxsize75mm\epsfbox{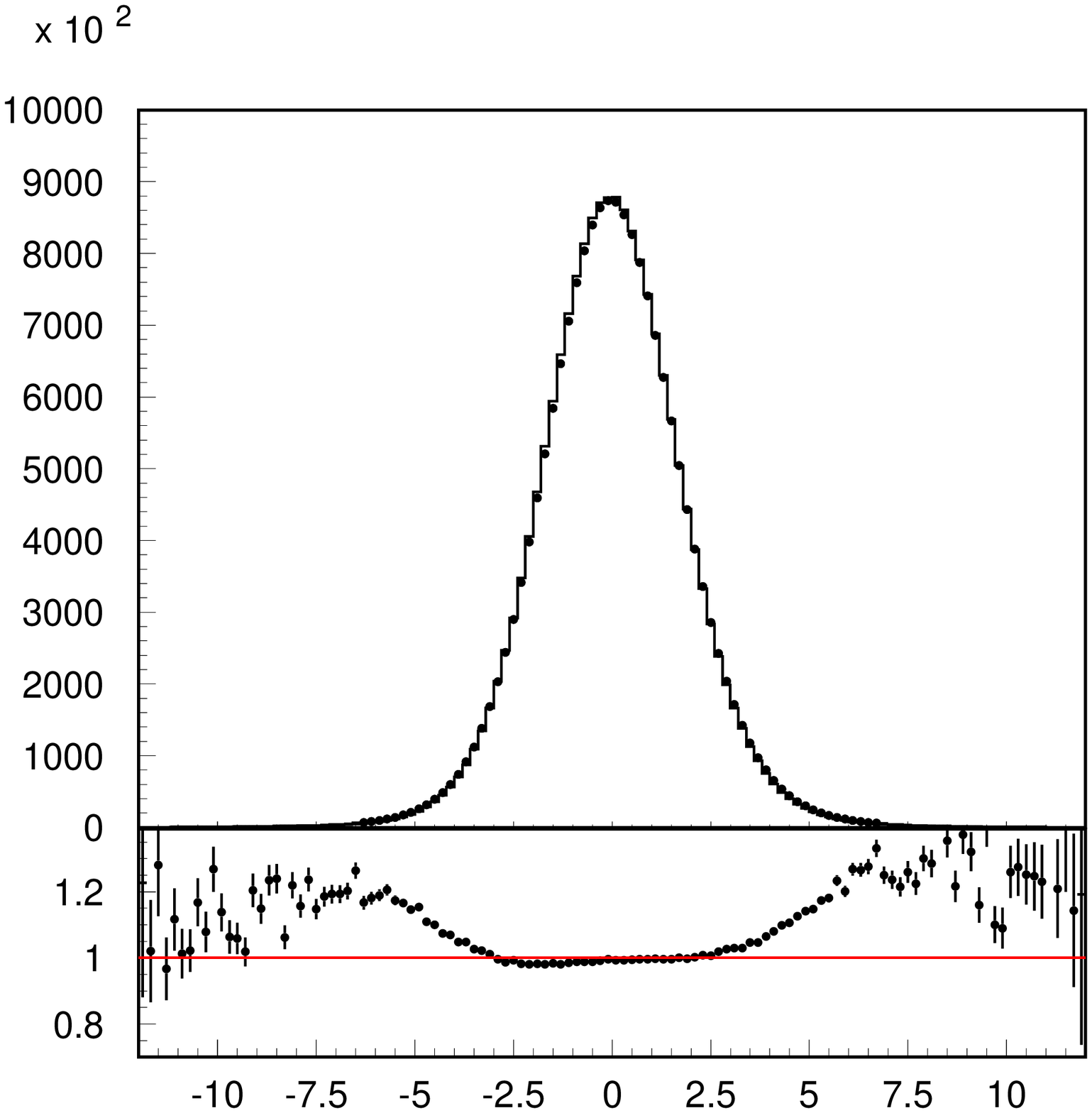}}
\put(100,-50){\epsfxsize75mm\epsfbox{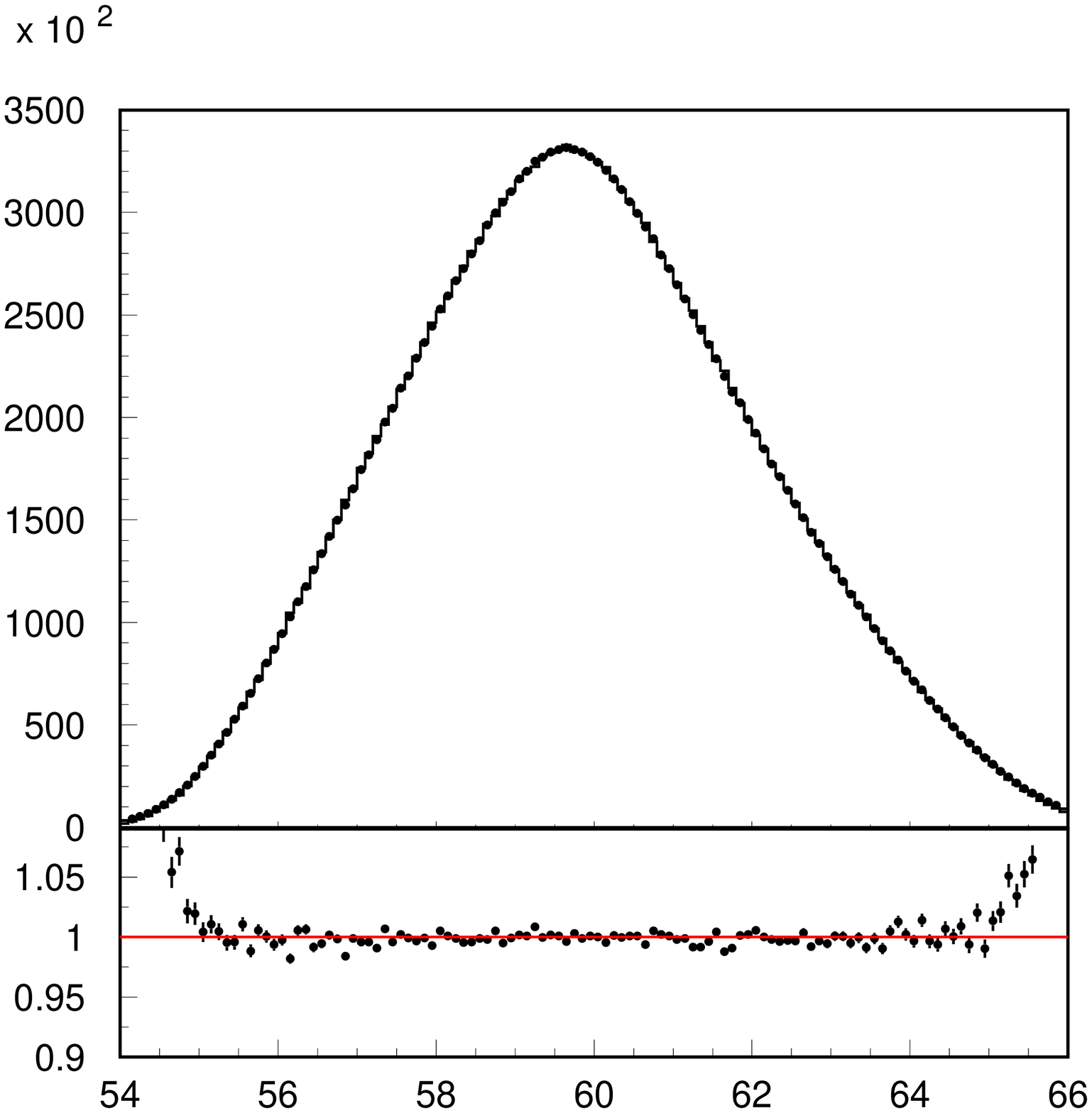}}
\put(-35,-50){\bf $M_{3\pi} - M_K  ~~(\MEVcc)$}
\put(215,-50){\bf $P_{3\pi} ~~(\GEVc)$}
\put(47,127){\bf \large (a)}
\put(272,127){\bf \large (b)}
\end{picture}
\end{center}
\vspace{5mm}
\caption{Distributions of the reconstructed $M_{3\pi}$ invariant mass {\bf (a)}
and the reconstructed kaon momentum {\bf (b)} for the normalization $\KTP$ 
candidates within the final selection. Data are shown as symbols and simulation
normalized to data as histograms. The lower plots are the ratios 
of the distributions (data/simulation) displayed in the upper plots.
\label{fig:norm}}
\end{figure}

The reconstructed $M_{3\pi}$ invariant mass spectrum is displayed in 
Fig.~\ref{fig:norm}a. Its measured resolution $\sigma_{3\pi}=1.7~\MEVcc$ is 
in agreement with simulation. The three track momentum sum distribution is 
shown in Fig.~\ref{fig:norm}b. The residual disagreement between data and 
simulation is considered in the systematic uncertainties study.

The number of prescaled $\KTP$ candidates in the signal region is 
$18.82 \times10^6 ~(12.09 \times10^6 ~\KPL$ and $6.73 \times10^6 ~\KMI$) with 
a negligible background.

%% file: ke4_accep.tex
A detailed GEANT3-based~\cite{geant} Monte Carlo (MC) simulation is used to 
compute the acceptances for signal and normalization channels.
It includes full detector geometry and material description,
stray magnetic fields, DCH local inefficiencies and misalignment, LKr local
inefficiencies, accurate simulation of the kaon beam line, and time variations 
of the above throughout the running period. This simulation is used to 
achieve a large time-weighted MC production, providing a simulated event
sample about 20 times larger than the signal sample and 1/4 of the prescaled 
normalization sample, reproducing the observed flux ratio 
$(\KPL / \KMI) \sim 1.8$.

The $\KEQ$ signal channel is generated according to the most precise 
description of the form factors as obtained in~\cite{ke410}. 
The normalization channel $\KTP$ is well understood in terms of simulation, 
being of primary physics interest to NA48/2~\cite{ba07}. The most precise 
values of the slopes of the Dalitz plot have been implemented~\cite{slopes}.
Attraction/repulsion between opposite charge/same charge particles and 
real photon emission using {\tt PHOTOS} 2.15~\cite{photos} are included 
in both simulations.

The same selection and reconstruction  as described in section~\ref{sec:selevt}
are applied to the simulated events except for the trigger and timing cuts.
Particle identification cuts related to the LKr response are replaced by
momentum-dependent efficiencies, obtained from pure samples of electron and
pion tracks.

The acceptances averaged over periods with different data taking conditions 
and over the two kaon charges are ($18.193 \pm 0.004) \%$  and 
$(23.967 \pm 0.010) \%$ for the signal and normalization channels, 
respectively. Due to the detector and beam line being largely charge symmetric 
by design, and due to the data taking conditions,
these values are practically identical for $\KPL$ and $\KMI$.
The uncertainty on the acceptance calculations due to the limited size of the
simulation samples (a few $10^{-4}$ relative) is included in the 
systematic error.

%% file: ke4_brsyst.tex
A large number of possible effects have been studied and quantified,
many of them being upper limits.
When necessary, a correction is applied to account for
any observed bias, and residual effects are quoted as systematic uncertainty. 
The considered contributions are described below.

{\bf Acceptance stability.} 
Many studies have been performed varying in turn the value of each cut applied 
in the common, signal and normalization selections. The maximum deviation 
observed with respect to the value of the reference cut has been quoted as the
uncertainty if statistically significant. 
None of the studied contributions are dominant and all are below the per mil 
relative level. Varying the common selection cuts contributes $0.03\%$ to the
relative systematic uncertainty of BR$(\KEQ)$. In the 
signal selection, the anti-background cuts amount to $0.03\%$ while the $\PK$
cut and residual momentum differences together contribute $0.08\%$.
The normalization selection cuts add another $0.08\%$. Momentum cuts in the 
particle identification contribute $0.05\%$ each when considering 
{\it electron} and {\it pion} definitions.
 
Time control of the beam geometry and acceptance modeling have been
investigated in detail. While the acceptances for signal and normalization 
show a relative variation of $\sim2$ percent between different data taking 
periods, related to beam geometry and DCH local inefficiencies,
their ratio is stable in time. A relative change of $0.07\%$ with respect to 
the nominal result is observed when simulated samples used to compute the 
acceptances are swapped between subsamples of the data before combining them. 
This value is assigned as an upper limit of the systematic uncertainty due to 
time variation of acceptance and beam geometry. It is fully consistent with the
variations observed when considering smaller subsamples of the data based on 
kaon beam charge and polarity of the achromat and spectrometer magnets. 

The impact of the limited precision of the measured relative form 
factors~\cite{ke410} on the signal acceptance has also been considered 
($0.06\%$). The modeling of the amount of material seen by decay particles 
before the magnet could affect bremsstrahlung emission of additional photons. 
As a result of the absence of explicit cuts on additional LKr activity from 
photons, the estimated $4\%$ precision on the simulated material thickness has 
only a $0.06\%$ impact on the final result. 

All the above uncertainties have been added in quadrature to a total relative 
contribution of $0.18\%$.

{\bf Muon vetoing efficiency.} The MUV veto requirement in the common selection
(Section~\ref{sec:selevt}) is essential in suppressing the $\KTP$ background to
an acceptable level\footnote{The background increases by a factor of 10 when 
this requirement is removed.}. 
Removing the MUV requirement in the selection of simulated signal and 
normalization events increases both acceptance values, but their ratio 
$A_n / A_s$ remains unchanged within $0.05\%$, suggesting that the 
rejection of late pion decays in flight does not bias the result. The 
probability to reconstruct a common three-track vertex decreases significantly 
when one or more pions decay to muons. The potential effect of the different 
number of final state pions in signal and normalization channels is therefore 
minimized
at the level of the common selection by requiring the presence of a good 
quality vertex. 
Stability of the result with respect to muon vetoing is also supported by 
varying the minimum track momentum cut from the nominal $5~\GEVc$ up to 
$10~\GEVc$, where the efficiency of MUV hit reconstruction varies by more than 
$10\%$ while the relative change of the final result is within $0.05\%$.

In order to estimate the potential bias from MUV reconstruction algorithm and 
combinatorial effects, the requirement for MUV hit association in the common 
selection has been modified to reject events with hits in all three rather than
in at least two MUV planes. The observed $0.16\%$ relative difference is 
conservatively quoted as a systematic uncertainty.

{\bf Accidental activity.}  
Possible accidental activity, either from beam particles or from fake
tracks ({\it ghost} tracks) resulting from DCH hit combinatorics\footnote{A 
{\it ghost} track is close to another track by less than 1 cm at the DCH1 
plane and is reconstructed with worse quality.}, has been subjected to a 
dedicated study.

\begin{figure}[htp]
\begin{center}
\begin{picture}(160,150)
\put(-155,-50){\epsfxsize75mm\epsfbox{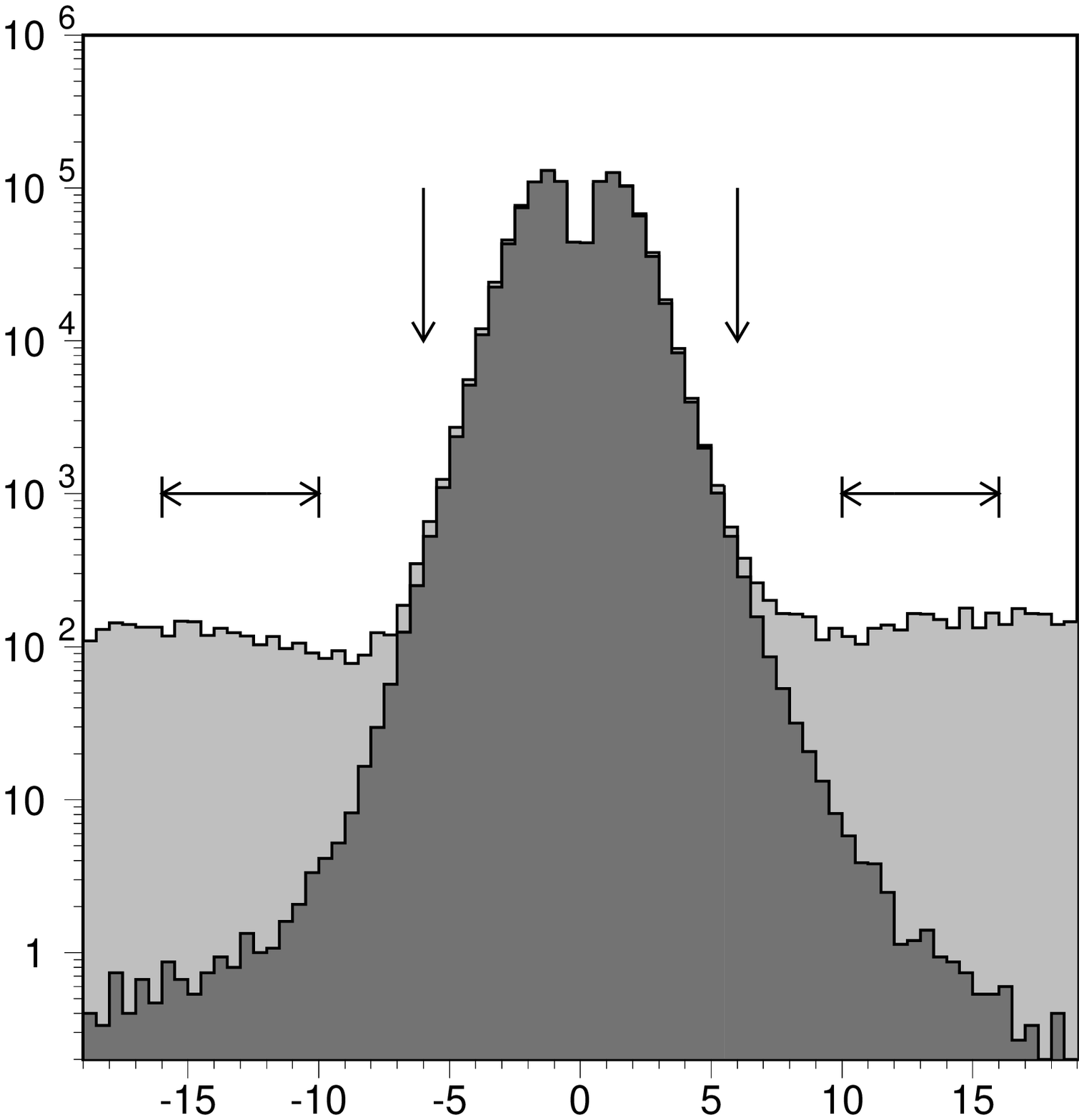}}
\put(90,-50){\epsfxsize75mm\epsfbox{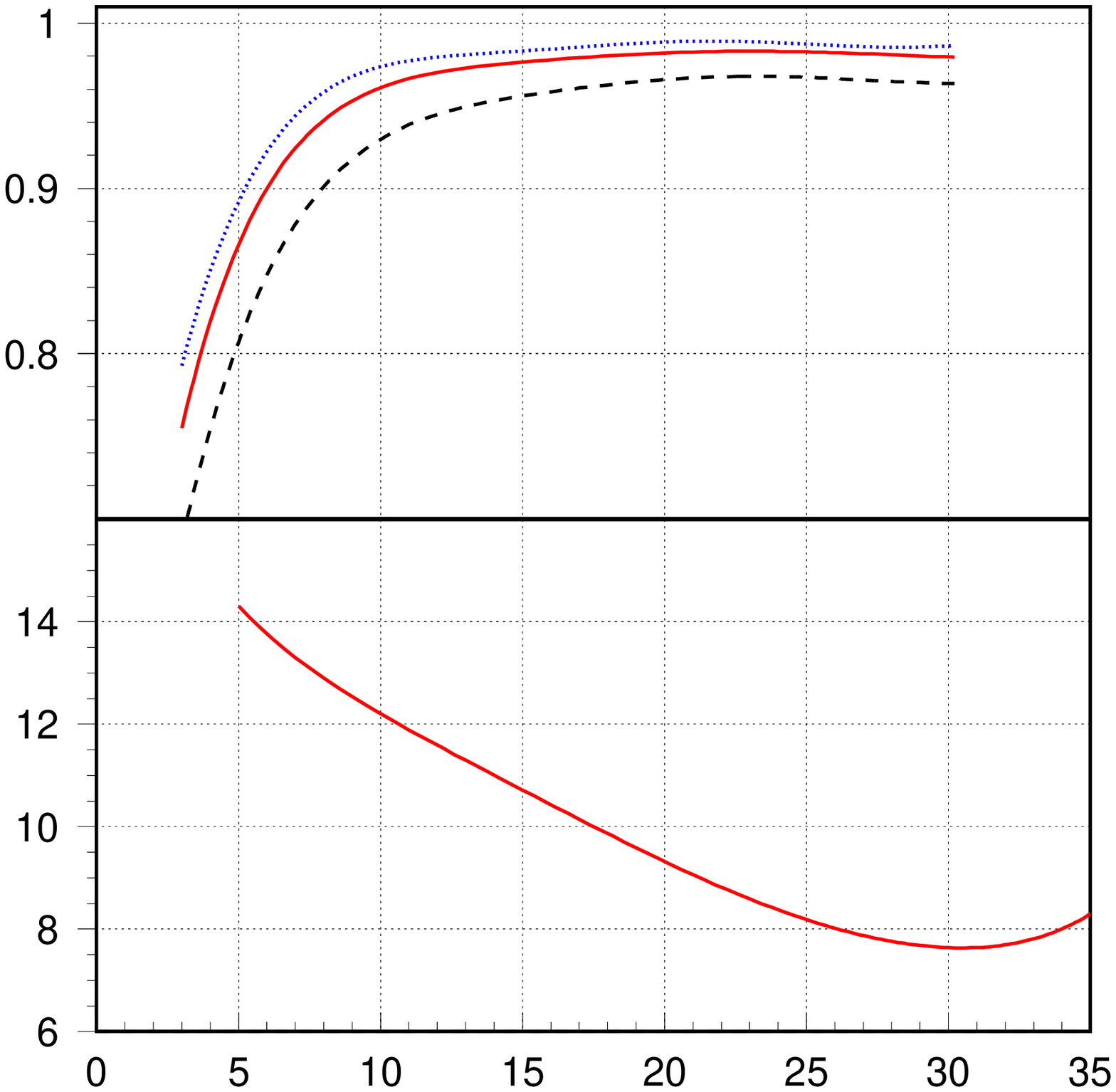}}
\put(-123,126){\bf \large (a)}
\put(120,126){\bf \large (b)}
\put(-121,70){\bf control}
\put(-6,70){\bf control}
\put(-60,125){\bf signal}
\put(-5,-50){$\Delta t$ (ns)}
\put(225,-50){$p ~~(\GEVc)$}
\put(205,85){\bf efficiency}
\put(190,70){\bf (electron track)}
\put(205,40){\bf rejection}
\put(195,25){\bf (pion track)}
\end{picture}
\end{center}
\vspace{5mm}
\caption{{\bf (a)} Normalized distribution of the time difference 
between the time of each vertex track and their average for $\KEQ$ (light 
shaded area) and $\KTP$ (dark shaded area) candidates. The vertical arrows 
indicate the signal time window. The horizontal arrows indicate the two control
regions. Only the time difference of the track with the largest $\Delta t$ 
absolute value is plotted. 
{\bf (b)} Efficiency of the linear discriminant variable as a function of
momentum for electron tracks (top curves). The solid line corresponds to the 
cut value of 0.90 applied in this analysis. The lower dashed line corresponds 
to the cut value of 0.95 and the upper dotted line to the cut value of 0.85. 
For illustration, the pion track rejection (bottom curve) is displayed for the 
cut value of 0.90.
\label{fig:syst}}
\end{figure}

The difference $\Delta t$ between the time of each vertex track and their 
average (the vertex time) is required  to be within $\pm6$~ns in the common 
selection. Removing this requirement reveals different tails in $\Delta t$
distribution for $\KTP$ and $\KEQ$ selections suggesting different 
contributions of accidental tracks forming a good vertex. 
The bias due to this effect is estimated by extrapolation from the control 
regions (--16; --10) ns and (+10; +16) ns to the nominal time window 
(accounting for WS events) as illustrated in Fig.~\ref{fig:syst}a. 
The subtraction of accidental background leads to a $-0.12\%$ correction to the
result, and the difference between two estimates (based on constant and 
parabolic extrapolation to the central signal window) is quoted as its 
uncertainty ($0.06\%$).

An event is rejected if an extra {\it non-ghost} track is present within a
6 ns window around the three-track vertex time.
A conservative estimate of the uncertainty due to the presence of accidental 
tracks not forming the decay vertex is obtained by variation of the above time 
limit up to 35 ns, and it is found to be $0.21\%$.

{\it Ghost} tracks are not allowed to form an accepted vertex and their 
presence in addition to the considered tracks is ignored in order to avoid bias
in $\KTP$ and $\KEQ$ samples related to the different reconstruction 
probabilities of fake tracks from pions and electrons. 
The {\it ghost} track tagging procedure using modified criteria 
(distance between tracks and quality of track reconstruction) has been studied 
in detail to identify the optimal one, and the residual systematic bias is 
estimated to be $0.04\%$.

{\bf Particle identification.} 
Different {\it pion} identification requirements have been studied, relaxing 
the $E/p$ condition and recomputing signal and normalization acceptances 
as described in section~\ref{sec:accep}. The largest difference to the
reference result $(0.08\%)$ is quoted as the related uncertainty. For the 
{\it electron} identification, varying the cut on the linear discriminant 
variable value between 0.85 and 0.95 (or even removing the cut) 
changes drastically the background contamination (up to a factor of four). 
Applying the corresponding momentum-dependent efficiency 
(Fig.~\ref{fig:syst}b) to the simulation, no bias is observed and a maximum 
deviation of $0.04\%$ from the reference result (obtained for the cut value of 
0.90) is observed. The uncertainties from {\it pion} and {\it electron} 
identification are added quadratically.

{\bf Background estimate.} The uncertainty on the scaling factor of 2 
used to estimate the background based on WS $\KEQ$ candidates has been studied 
in two ways. In the WS event selection, the square invariant mass $M_{X}^{2}$ 
(Section~\ref{sec:selevt}) is computed for both pions and can be used to 
classify further the origin of the event. If the smaller mass squared is above 
$(0.27~\GEVcc)^2$ (corresponding to $2 m_{\pi^+}$ with resolution smearing), 
the event is assigned a factor two weight as being $\KTP$-like $(95.8\%)$, 
otherwise it is assigned a factor one weight as being $\KPPD$-like $(4.2\%)$. 
This rough estimate leads to a factor 1.96.
Another estimate, based on a simulated $\KTP$ sample properly weighted for 
particle-identification performances, gives a similar ratio 
Right Sign electron/Wrong Sign electron =
($1.94 \pm 0.15$), confirming the prescription for WS background related to 
pion misidentification. It is also in agreement with the overall
factor of ($2.0 \pm 0.3$) used in the form factor analysis \cite{ke410}. 
The $\pm 0.15$ uncertainty is propagated to the result as a relative 
uncertainty of $0.07\%$, based on the above studies.  
In addition, variation of the background-related requirements in the 
$\KEQ$ selection within wide ranges shows excellent stability of the result.

{\bf Other sources.}
Dedicated MC samples simulated without real photon 
emission were used to estimate the impact from radiative effects description. 
One tenth of the full effect was assigned as a modeling uncertainty according 
to the prescription of \cite{ke4rad}.

Trigger efficiency accuracy is limited by the size of the
control samples. The statistical precision is quoted as a contribution to the 
systematic uncertainty.

Simulated samples used in the acceptance calculations contribute to the 
systematic uncertainty through their statistical precision: this source could
be reduced by increasing the simulation statistics but already contributes 
at a very low level.

Sizable uncertainty arises from the external input BR$(\KTP)$
known experimentally with a limited relative precision of $0.72\%$~\cite{pdg}.

Two independent analyses have been compared on a subsample of the data and 
found to be fully consistent, ensuring the robustness of the procedure.

\vspace{2mm}
\noindent The breakdown of the considered systematic uncertainties is 
displayed in Table~\ref{tab:brsys}. 

\begin{table}[htp]
\vspace{-3mm}
\begin{center}
\caption{Summary of the relative corrections applied to the BR$(\KEQ)$ value 
and relative contributions to the systematic uncertainty. 
\label{tab:brsys}}
\vspace{1mm}
\begin{tabular}{lcc}
\hline
Source              &  Correction  $(\%)$ & Contribution $(\%)$\\
                    &   to BR value       & to BR uncertainty \\
\hline
\multicolumn{3}{l}{\bf Common to all subsamples} \\
Acceptance stability & --     & 0.18 \\
Muon vetoing efficiency      & --     & 0.16 \\
Accidental activity          & --0.12 & 0.21 \\
Particle identification      & --     & 0.09 \\
Background estimate          & --     & 0.07 \\
Radiative events modeling    & --     & 0.08 \\
\multicolumn{3}{l}{\bf Subsample-dependent quoted as a global equivalent}  \\
Trigger efficiency           & --     & 0.11 \\ 
Simulation statistics        & --     & 0.05 \\
\hline
Total  systematics           & --0.12 & 0.37 \\
External error               & --     & 0.72 \\
\hline
\end{tabular}
\vspace{-1cm}
\end{center}
\end{table}

%% file: ke4_brresult.tex
\noindent The final result is a weighted average of 16 values
obtained in eight independent data subsamples and both kaon charges.
The weight of each input includes error contributions of time-dependent
statistical origin: event statistics (signal, background and normalization), 
trigger efficiencies and acceptances.
This method is more robust against time-dependent conditions than using an
averaged acceptance and trigger efficiency over the whole data taking period.
However, due to the careful time-dependent treatment of simulated samples, 
this potential difference is kept below the per mil relative level and
is taken into account in the systematic error. Other systematic uncertainties 
(Table~\ref{tab:brsys}) are common to all subsamples and are then quoted as a 
single error on the final result. All input ingredients to the BR$(\KEQ)$ 
measurement are summarized in Table~\ref{tab:brin}.
By convention, the uncertainties are assigned to three categories:
(i) statistical errors from the numbers of $\KEQ$ signal candidates (dominant 
error), WS data events (used for background computation) and normalization events;
(ii) subsample-dependent systematic uncertainties such as those of trigger 
efficiencies and acceptance 
and systematic uncertainties common to all subsamples;
(iii) the external error related to the uncertainty on the normalization mode
branching ratio (BR$($\KTP$) = (5.59 \pm 0.04)\%$~\cite{pdg}).

\begin{figure}[htp]
\begin{center}
\begin{picture}(160,240)
\put(-125,230){ \large  BR($\KEQ) \times 10^5$ }
\put(195,35){ \large Sample }
\put(-140,-140){\epsfxsize150mm\epsfbox{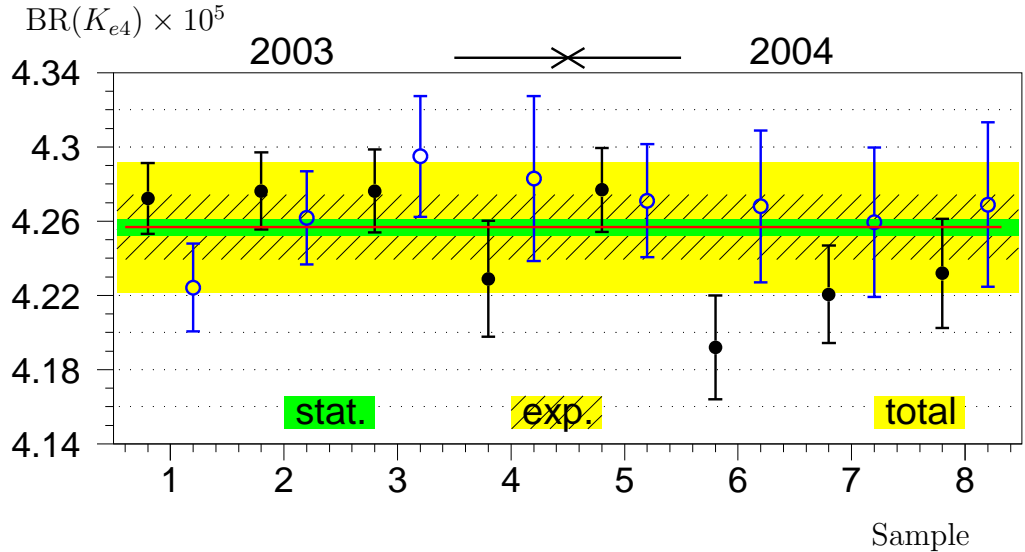}}
\end{picture}
\end{center}
\vspace{-25mm}
\caption{$\KEQ$ branching ratio for eight statistically independent samples and
both kaon charges. The hatched band shows the experimental error 
$(\sigma\expe = \sigma\stat \oplus  \sigma\syst)$. The total error 
(shaded band) includes the external error. 
The fit $\chi^2$ is 15.85 for 15 degrees of freedom when including the 
time-dependent errors only. Full symbols correspond to $\KPL$ results and empty
symbols to $\KMI$ results. 
\label{fig:br}}
\end{figure}

The resulting values, including all errors, are found to be:
\begin{eqnarray}\label{eq:ke4rate}
\Gamma(\KEQ) / \Gamma(\KTP) = (7.615 \pm 0.008\stat \pm 0.028\syst) \times 10^{-4}
\end{eqnarray}
and
\be
{\rm BR}(\KEQ) = (4.257 \pm0.004\stat \pm0.016\syst \pm0.031\ext)\times10^{-5},
\label{eq:result}
\ee
where the branching ratio error is dominated by the external uncertainty from 
the normalization mode. The BR$(\KEQ)$ values obtained for the statistically 
independent subsamples are shown in Fig.~\ref{fig:br}, also in perfect 
agreement with the global value obtained from the whole sample and the values
measured separately for $\KPL$ and $\KMI$:
\begin{displaymath}
{\rm BR}(\KEQP) = (4.255\pm0.008)\times10^{-5},~~~
{\rm BR}(\KEQM) = (4.261\pm0.011)\times10^{-5}, 
\end{displaymath}
where the quoted uncertainties include statistical and time-dependent
systematic contributions.

\begin{table}[htp]
\begin{center}
\caption{Inputs to the BR$(\KEQ)$ measurement for $\KPL,\KMI$ and combined $\KPM$.
The relative contribution of each item to the BR($\KEQPM$) statistical 
uncertainty is also shown in the last column. Statistical errors on the 
acceptance and trigger efficiency values (given within parentheses) are
taken into account in the systematic error (Table~\ref{tab:brsys}) and not in 
the total statistical error given in the last row.
\label{tab:brin}}
\vspace{1mm}
\begin{tabular}{lrrr|c}
\hline
                    & $\KPL$  & $\KMI$  & $\KPM$ & BR$(\KEQPM)$  \\
          &            &           &          &  relative error $(\%)$ \\
\hline
Signal events             &    712 288 &   396 653 &  1 108 941 & 0.096 \\
WS events                 &      3 276 &     2 000 &      5 276 & 0.013 \\
Normalization events/100  & 12 090 376 & 6 728 544 & 18 818 920 & 0.023 \\
$\KEQ$ acceptance ($\%$)  & 18.190     & 18.197    & 18.193     & (0.020) \\
$\KTP$ acceptance ($\%$)  & 23.970     & 23.961    & 23.967     & (0.041) \\ 
$\KEQ$ trigger efficiency ($\%$) &  98.546 & 98.480 & 98.523 & (0.108) \\ 
$\KTP$ trigger efficiency ($\%$) &  97.634 & 97.687 & 97.653 & (0.033) \\ 
\hline
\multicolumn{4}{l|}{Total relative statistical error $(\%)$} & 0.100 \\
\hline
\end{tabular}
\vspace{-10mm}
\end{center}
\end{table}

%% file: ke4_fftheo.tex
\begin{sloppypar}
The $\KEQ$ decay rate (in s$^{-1}$) is described in the five-dimensional space 
of the Cabibbo--Maksymowicz kinematic variables~\cite{cabmak}, namely the 
dipion $(\SP)$ and dilepton $(\SE)$ squared invariant masses and the three 
decay angles [$\theta_{\pi}(\TE)$, the same sign pion (electron) angle in 
the dipion (dilepton) rest frame to the dipion (dilepton) line of flight in the
kaon rest frame, and $\phi$, the angle between the dipion and dilepton planes 
in the kaon rest frame] as:
\end{sloppypar}

\be\label{eq:dkp}
  \\
d\Gamma_{5} = 
 \displaystyle\frac{G_{F}^{2}  |V_{us}|^{2}}{2 \hbar (4 \pi)^{6} m_{K}^5} ~\rho(\SP,\SE) ~J_{5}(\SP,\SE,\CTP,\CTE,\phi)
  ~d \SP ~d \SE ~d \CTP ~d\CTE ~d\phi,  \\
  \\
\ee

\noindent where $ \rho(\SP,\SE) = X \sigma_{\pi}\left( 1 - \ZE \right)$  
is the phase space factor, with
 $
  ~X = \frac{1}{2} \lambda^{1/2}(m_{K}^2,\SP,\SE),
  ~\sigma_{\pi} = (1 -4m _{\pi}^{2}/\SP)^{1/2},
  ~\ZE = \MES / \SE$ and
$\lambda(a,b,c) = a^2 + b^2 + c^2 - 2 ( ab + ac + bc). $
The function $J_{5}$, using four combinations of $F,~G,~R,~H$ complex hadronic
form factors ($F_{i}, i=1,4$), reads~\cite{paist}:

\[
\begin{array}{ll}
   &     \\

J_{5} = 2 (1 - \ZE) & (I_{1} + I_{2}~\CTTE + I_{3}~\SSTE \cdot\cos2\phi
 + I_{4}~\STTE \cdot \CPH  + I_{5} ~\STE \cdot\CPH  \\
                          & + I_{6}~\CTE + I_{7}~\STE\cdot\SPH
 + I_{8}~\STTE \cdot\SPH + I_{9}~\SSTE \cdot\sin2\phi), \\

\end{array}
\]
where
\[
\begin{array}{rl}
I_{1} &= \QRT~\left( (1 + \ZE )|F_{1}|^{2} + \HLF(3+ \ZE)(|F_{2}|^{2} + |F_{3}|^{2})~\SSTP + 2 \ZE |F_{4}|^{2}\right) , \vspace{2mm}\\
I_{2} &= -\QRT ~(1 - \ZE)\left( |F_{1}|^{2} -\HLF (|F_{2}|^{2} + |F_{3}|^{2})~\SSTP \right), \vspace{2mm}\\
I_{3} &= -\QRT ~(1 - \ZE)\left( |F_{2}|^{2} - |F_{3}|^{2} \right)~\SSTP  ,\vspace{2mm}\\
I_{4} &= \HLF ~(1 - \ZE) Re(F_{1}^{*}F_{2})~\STP,\vspace{2mm}\\
I_{5} &= - ~\left(Re(F_{1}^{*}F_{3}) + \ZE~Re(F_{4}^{*}F_{2}) \right)~\STP, \vspace{2mm}\\
I_{6} &= - ~\left(Re(F_{2}^{*}F_{3})~\SSTP - \ZE~Re(F_{1}^{*}F_{4}) \right),\vspace{2mm} \\
I_{7} &= - ~\left(Im(F_{1}^{*}F_{2}) + \ZE~Im(F_{4}^{*}F_{3}) \right)~\STP, \vspace{2mm} \\
I_{8} &= \HLF~(1 - \ZE) ~Im(F_{1}^{*}F_{3} )~\STP,\vspace{2mm}\\
I_{9} &= -\HLF(1 - \ZE)~ Im(F_{2}^{*}F_{3})~\SSTP .
\end{array}
\]

In $\KEQ$ decays, the electron mass can be neglected $(\ZE = 0)$ and the
terms $( 1 \pm \ZE )$ become unity. One should also note that the form
factor $F_{4}$ is always multiplied by $\ZE$ and thus does not contribute
to the full expression\footnote{The form factor $R$ enters only in the 
definition of $F_4$ and therefore cannot be addressed in the $\KEQ$ decay 
analysis}.
With this simplification, the complex hadronic form factors $F_{i}$ reduce to:

\be
F_{1} = m_{K}^2 (\gamma~F + \alpha~G~\CTP), ~~~F_{2} = m_{K}^2 (\beta~G), ~~~F_{3} = m_{K}^2 (\beta \gamma~H),
\label{eq:f123}
\ee

\noindent where one uses the three dimensionless complex form factors $F,G$ (axial),
$H$ (vector), and three dimensionless combinations of the $\SP$ and $\SE$
invariants:
\[
\begin{array}{lll}
\alpha = \sigma_{\pi} (m_{K}^2-\SP-\SE)/2 m_{K}^2 , &~\beta  =  \sigma_{\pi} (\SP\SE)^{1/2}/ m_{K}^2 , &~\gamma =  X / m_{K}^2 .
\end{array}
\]

A further partial wave expansion of the form factors $F_{1,2,3}$ with respect 
to the $\CTP$ variable is considered~\cite{paist} and, once limited to S- and 
P-wave terms and assuming the same phase $\delta_{p}$ for all P-wave form 
factors, leads to the expressions for $F,G,H$:
\begin{eqnarray}\label{eq:spwav}
 F & = & F_{s} ~e^{i\delta_{s}} + F_{p} ~e^{i\delta_{p}}\CTP ~,  \nonumber \\
 G & = & G_{p} ~e^{i\delta_{p}} ~,   \\
 H & = & H_{p} ~e^{i\delta_{p}} ~. \nonumber
\end{eqnarray}

The model-independent analysis in~\cite{ke410} determines simultaneously the 
four real numbers $F_{s}$, $F_{p}$, $G_{p}$, $H_{p}$ and the phase difference 
$(\delta =\delta_{s} - \delta_{p})$ in bins of $\SP, \SE$. 

In presence of electromagnetic interaction, the differential decay 
rate~(\ref{eq:dkp}) 
is modified by the presence of virtual and real photon emission. This effect is
implemented in two steps. First, the Coulomb attraction/repulsion between two 
opposite/same charge particles is considered:
\[C(S_{ij}) = \prod_{i \neq j} \frac{\omega_{ij}}{e^{\omega_{ij}}-1} \]
with $\omega_{ij} = 2\pi\alpha Q_i Q_j /\beta_{ij}$, where
$\alpha$ is the fine structure constant, $Q_i Q_j = -1$ for opposite charge
particles ($+1$ for same charge particles) and $\beta_{ij}$ is their
relative velocity (in units of $c$). 
The largest effect comes from the attraction between the two pions at
low relative velocity. It depends only on the $\SP$ variable. The electron
(positron) being relativistic, the other attractive/repulsive pion-electron 
terms are constant and their product amounts to $0.9998$. 
Then, the {\tt PHOTOS} 2.15 program~\cite{photos} interfaced
to the simulation is used for real photon emission. Its effect is a distortion
of the kinematic variable distributions and is evaluated on a grid of the 
5-dimensional space.

%% file: ke4_ff.tex
\subsection{Form factor determination}

The $\KEQ$ decay form factors were extensively studied~\cite{ke410} with 
the same data sample as used for the present analysis and 
their energy variation described as a series expansion of the dimensionless 
invariants $q^{2} = (\SP/4m_{\pi}^{2}) -1$ and $\SE/4m_{\pi}^{2}$. All values
have been given relative to a common value $f_s$, the S-wave axial vector
form factor $F_s(q^{2} = 0 , \SE = 0)$ :  

\begin{eqnarray}\label{eq:fexp}
F_{s}/f_{s} & = &1 + f^{\prime}_{s}/f_{s}~q^{2} + f^{\prime\prime}_{s}/f_{s}~q^{4} + f^{\prime}_{e}/f_{s}~\SE/4m_{\pi}^{2}~, \nonumber \\
F_{p}/f_{s} & = &f_{p}/f_{s}~,  \nonumber \\
G_{p}/f_{s} & = &g_{p}/f_{s} + g^{\prime}_{p}/f_{s} ~q^{2}, \nonumber \\
H_{p}/f_{s} & = &h_{p}/f_{s}~.
\end{eqnarray}

\noindent  Integrating $d\Gamma_{5}$  (\ref{eq:dkp}) over the 5-dimensional
space after substituting $F_1 ,F_2, F_3$ (\ref{eq:f123}) by their 
expression and measured values (\ref{eq:spwav}, \ref{eq:fexp}), including 
radiative effects and leaving out the $|V_{us}|$ and $f_s$ constants, the 
$\KEQ$ branching ratio, inclusive of radiative decays, is obtained as:

\be\label{eq:ffbr}
{\rm BR}(\KEQ) = 
\tau_{K^{\pm}} \cdot (|V_{us}| \cdot  f_{s})^{2} \cdot \int d\Gamma_{5}~ / ~( |V_{us}| \cdot  f_{s})^{2} ,
\ee
where $\tau_{K^{\pm}}$ is the $\KPM$ mean lifetime (in seconds). The value of 
$f_{s}$ is then obtained from the measured value of BR$(\KEQ)$ and the 
integration result.

Because of the quadratic dependencies displayed in (\ref{eq:ffbr}), the
relative uncertainty on $f_{s}$ is only half the relative uncertainty from 
BR$(\KEQ)$, kaon lifetime and kinematic space integral, while any relative 
uncertainty on $V_{us}$ propagates with full size. Contributions are 
categorized as follows:\\
--- Statistical error stems only from the BR$(\KEQ)$ measurement. \\
--- Systematic uncertainties originate from the BR$(\KEQ)$ measurement
and the phase space integral evaluation.
Uncertainties on the integration result when varying each relative form factor
and energy dependence within $\pm 1 \sigma$  have been considered. The known 
large anti-correlations between $f^{\prime}_{s},f^{\prime\prime}_{s}$  and
$g_{p},g^{\prime}_{p}$ have been omitted to be conservative. 
The detailed description of the phase shift between S- and P-wave form factors 
(\ref{eq:spwav}) has a negligible impact.
The robustness of the integration method has also been checked against several
integration grid definitions. 
As in the branching ratio and relative form factor  measurements, one tenth of 
the full {\tt PHOTOS} effect is assigned as systematic uncertainty on the 
radiative corrections modeling. \\
--- External inputs contributing to the  $f_{s}$ form factor uncertainty are 
related to the kaon lifetime $\tau_{K^{\pm}}$, 
the branching ratio of the normalization decay mode BR$(\KTP)$ and the 
$|V_{us}|$ value. All quantities are taken from~\cite{pdg}.
However, it should be kept in mind that only the product $|V_{us}| \cdot f_s$ 
is accessible by this measurement. 

Table~\ref{tab:ffsyst} summarizes the error contributions.

\begin{table}[htp]
\begin{center}
\caption{Summary of the contributions to the $f_s$ form factor uncertainties.
\label{tab:ffsyst}}
\vspace{1mm}
\begin{tabular}{lc}
\hline
Source &  relative contribution $(\%)$\\
\hline
BR$(\KEQ)$ statistical error & 0.05 \\
BR$(\KEQ)$ systematic error &  0.20 \\
Form factor energy dependence (systematic error) &  0.21 \\
Integration method (systematic error) &  0.02\\
Radiative effects in integration (systematic error) & 0.04 \\
\hline
Total experimental error &  0.30   \\
\hline
BR$(\KEQ)$ external error & 0.36 \\
Kaon lifetime (external error) & 0.08 \\
$|V_{us}|$  (external error) & 0.40 \\
\hline
Total error (including external errors) & 0.62 \\
\hline
\end{tabular}
\vspace{-5mm}
\end{center}
\end{table}

%% file: ke4_ffresult.tex
\noindent Given the measured $\KEQ$ branching ratio value (\ref{eq:result}) and
using the world average kaon lifetime value $(1.2380 \pm 0.0021) \times 10^{-8}$ s,
the measurement of the form factors~\cite{ke410} is now complemented by the 
overall $f_{s}$ normalization:
\begin{eqnarray} 
|V_{us}| \cdot f_s = & 1.285 \pm 0.001  \stat \pm 0.004 \syst \pm 0.005 \ext \\
{\rm corresponding~to~} f_s = & 5.705 \pm 0.003 \stat \pm 0.017 \syst \pm 0.031 \ext
\end{eqnarray}

\noindent when using $|V_{us}|=0.2252 \pm 0.0009$~\cite{pdg}. 

The obtained $f_s$ value and its error can be propagated to all relative form 
factors now displayed with absolute values in Table~\ref{tab:absff} and 
including an additional normalization error, fully correlated over all 
measured values.

\begin{table}[htb]
\begin{center}
\caption{
\label{tab:absff} Absolute values of the  form factor measurements (as defined 
in (\ref{eq:fexp})).
There are large anti-correlations between $f^{\prime}_{s} , f^{\prime\prime}_{s} ~(-0.954)$ and $g_p ,g^{\prime}_p  ~(-0.914)$. 
The normalization error is fully correlated over all form factors.}
\vspace{2mm}
\begin{tabular}{ll}
\hline
$f_{s}$ & $= \phantom{-}5.705 \pm~0.003 \stat \pm~0.017 \syst \pm~0.031 \ext$ \\
$f_{s}$ & $= \phantom{-}5.705 \pm~0.035\com$  \\
\hline
$f^{\prime}_{s}$ & $= \phantom{-}0.867 \pm~0.040\stat\pm~0.029\syst \pm~0.005\com$ \\
$f^{\prime\prime}_{s}$& $= -0.416 \pm~0.040\stat \pm~0.034\syst  \pm~0.003\com$ \\
$f^{\prime}_{e}$& $= \phantom{-}0.388 \pm~0.034\stat \pm~0.040\syst \pm~0.002\com$\\
$f_p$& $=  -0.274 \pm~0.017\stat \pm~0.023\syst  \pm~0.002\com$ \\
$g_p $ & $= \phantom{-}4.952 \pm~0.057\stat \pm~0.057\syst \pm~0.031\com$\\
$g^{\prime}_p$& $ = \phantom{-}0.508 \pm~0.097\stat \pm~0.074\syst \pm~0.003\com$ \\
$h_p$& $ = -2.271 \pm~0.086\stat \pm~0.046\syst \pm~0.014\com$ \\
\hline
\end{tabular}
\vspace{-5mm}
\end{center}
\end{table}

In addition to the above set of values, it can be of  further
theoretical interest to quote also the S- and P-wave normalized 
projections of the $F_1$ form factor:

\be\label{eq:f1p}
F_{1} / \gamma m_{K}^2 = F_s ~e^{i\delta_{s}} + (F_p + \alpha/\gamma ~G_p )~\CTP ~e^{i\delta_{p}}, 
\ee
namely $ F_s$ and $(F_p + \alpha/\gamma ~G_p)$, respectively.
As all form factors are obtained in simultaneous fits together with the phase
difference $\delta_{s} - \delta_{p}$~\cite{ke410}, they exhibit correlations 
which vary with energy. In particular, the fit parameters $F_p$ and $G_p$ are 
strongly anti-correlated with a coefficient close to unity. The combination 
$\tilde{G_p} = G_p + \gamma/\alpha ~F_p$ shows much less correlation with 
$G_p ~(\sim0.20$ at most) and is also obtained in the fit. To allow an easy 
interpretation of the results without the explicit description of the fit 
correlations, the values of $F_s ,~\tilde{G_p}$ are given in Table~\ref{tab:fg}
together with those of $F_p ,G_p$ and $H_p$. Using (\ref{eq:f1p}), the P-wave
$F_1$ normalized projection can then be obtained as 
$\alpha/\gamma ~\tilde{G_p}$. It can be noted that for $\SE = 0$, the factor
$\alpha/\gamma$ reduces to $\sigma_{\pi}$.

\begin{table}[htb]
\begin{center}
\caption{\label{tab:fg} Absolute values of form factor measurements in ten 
$\MPP$ bins. First error within parentheses is statistical, second is 
systematic (bin to bin uncorrelated part only). A common relative error of 
$0.62\%$ must be added to each form factor bin by bin measurement, fully 
correlated over all form factor and bin measurements.
$ F_s$ values correspond to the projection of $F_s (\SP,\SE)$ on the $\MPP$ 
axis. No significant $\SE$ dependence has been observed for 
$F_p ,~G_p , ~\tilde{G_p}$ and $H_p$ within the available statistics.}
\vspace{1mm}
\begin{tabular}{cccc}
\hline
 Bin   & $\MPP$ barycenter &  \multicolumn{2}{c}{dimensionless form factors}\\
number&  $(\MEVcc)$      & $F_{s}$ & $\tilde{G_p}$  \\
\hline
1  & 286.06 & 5.7195(3)(3) & 4.334(74)(19)\\
2  & 295.95 & 5.8123(3)(1) & 4.422(53)(31)\\
3  & 304.88 & 5.8647(3)(2) & 4.550(46)(25)\\
4  & 313.48 & 5.9134(3)(2) & 4.645(41)(23)\\
5  & 322.02 & 5.9496(3)(1) & 4.711(38)(28)\\
6  & 330.80 & 5.9769(3)(1) & 4.767(35)(27)\\
7  & 340.17 & 6.0119(3)(1) & 4.780(34)(30)\\
8  & 350.94 & 6.0354(3)(1) & 4.907(34)(20)\\
9  & 364.57 & 6.0532(3)(1) & 5.019(35)(19)\\
10 & 389.95 & 6.1314(3)(5) & 5.163(36)(21)
\end{tabular}
\begin{tabular}{cccc}
\hline
 Bin   &   \multicolumn{3}{c}{dimensionless form factors}\\
number &   $F_{p}$       &     $G_{p}$      & $H_{p}$\\
\hline
1  & $-0.181(67)(15)$ & 5.053(258)(66)& $-1.795(518)(193) $\\
2  & $-0.324(62)(34)$ & 5.186(142)(84)& $-2.088(320)(~77) $\\
3  & $-0.209(60)(33)$ & 4.941(108)(59)& $-1.995(267)(~98) $\\
4  & $-0.156(58)(32)$ & 4.896(~91)(51)& $-2.750(246)(~72) $\\
5  & $-0.366(55)(41)$ & 5.245(~80)(58)& $-2.045(237)(~98) $\\
6  & $-0.383(54)(38)$ & 5.283(~73)(56)& $-2.705(234)(~88) $\\
7  & $-0.218(55)(46)$ & 5.054(~68)(59)& $-2.203(235)(156) $\\
8  & $-0.302(54)(33)$ & 5.264(~62)(37)& $-1.856(239)(110) $\\
9  & $-0.309(54)(31)$ & 5.357(~57)(30)& $-2.096(251)(217) $\\
10 & $-0.264(59)(33)$ & 5.418(~55)(33)& $-2.865(287)(177) $\\
\hline
\end{tabular}
\end{center}
\end{table}

%% file: ke4_conclu.tex
From a sample of $1.11 \times 10^{6}$ $\KEQ$ decay candidates with $0.95\%$ 
background contamination, the branching fraction, inclusive of $\KEQ_\gamma$
decays,  has been measured to be 
${\rm BR}(\KEQ) =(4.257\pm0.016\expe \pm0.031\ext) \times 10^{-5}$ 
using  $\KTP$ as normalization mode (the experimental error $\sigma\expe$ is 
the quadratic sum of the statistical $\sigma\stat$ and systematic $\sigma\syst$
uncertainties).
 The relative $0.8\%$ precision of the achieved measurement, dominated by the 
external uncertainty from the normalization mode, represents a factor of 
$\sim3$ improvement with respect to the world average value,
BR$(\KEQ) = (4.09 \pm0.10) \times 10^{-5}$ based on two earlier 
measurements~\cite{S118,E865}.
The relative decay rate 
$\Gamma(\KEQ) / \Gamma(\KTP) = (7.615 \pm 0.030\expe) \times 10^{-4}$ is 
measured with a $0.4\%$ relative precision, a factor of $\sim5$ improvement
over the current world average value of $(7.31 \pm 0.16) \times 10^{-4}$.

The hadronic form factors that characterize the decay have been evaluated
both for absolute value and energy dependence. The overall normalization
form factor $F_{s}(q^2 = 0, \SE = 0)$ has been measured with a $0.6\%$ total
relative precision as $f_s = 5.705 \pm 0.017 \expe \pm 0.031 \ext$ 
when using values of kaon mean lifetime $\tau_K$ and $|V_{us}|$ 
from ~\cite{pdg}, a factor of $\sim2$ and 4 improvement with 
respect to the values $f_s = 5.75 \pm 0.08$ ~\cite{E865} and 
$f_s = 5.59 \pm 0.14$ ~\cite{S118} obtained by earlier experiments.
The achieved improved precision on $\KEQ$ rate and form factors brings new
inputs to further theoretical studies and allows stronger tests of Chiral 
Perturbation Theory predictions.